\documentclass[reqno,11pt]{amsart}

\RequirePackage{lineno}
\usepackage{fancyhdr}
\usepackage[svgnames]{xcolor}
\usepackage{setspace}
\usepackage{pgf}
\usepackage{tikz}
\usepackage{mathpazo} 
\usepackage[scaled]{helvet} 
\usepackage{courier} 
\normalfont
\usepackage[T1]{fontenc}

\usetikzlibrary{patterns}
\usepackage{soul}
\usepackage{mathrsfs}
\usepackage{stmaryrd}
\usepackage[hidelinks=true]{hyperref}
\usepackage{natbib}
\usepackage{xfrac}
\usepackage[color=Orange]{todonotes}
\usepackage{layout}
\usepackage{microtype}
\usepackage{fancyhdr}
\usepackage{mathtools}

\usepackage{cleveref}
\crefname{figure}{figure}{figures}
\creflabelformat{figure}{#2#1#3}
\crefname{eq.}{equation}{equations}
\creflabelformat{equation}{#2(#1)#3}
\crefname{lemma}{lemma}{lemmas}
\creflabelformat{lemma}{#2#1#3}
\crefname{theorem}{theorem}{theorems}
\creflabelformat{lemma}{#2#1#3}
\crefname{condition}{condition}{conditions}
\creflabelformat{condition}{#2#1#3}
\crefname{assumption}{assumption}{assumptions}
\creflabelformat{assumption}{#2#1#3}
\crefname{appendix}{appendix}{appendices}
\creflabelformat{appendix}{#2#1#3}
\crefname{enumi}{}{}
\creflabelformat{enumi}{(#2#1#3)}

\newtheorem{theorem}{Theorem}

\newtheorem{lemma}{Lemma}
\newtheorem{definition}{Definition}

\newtheorem{assumption}{Assumption}

\usepackage{amssymb}

\DeclareMathOperator*{\argmax}{argmax}
\DeclareMathOperator*{\Det}{det}

\newcommand{\one}{\mathbb{1}}

\newcommand\abstraction{
This paper studies social interactions in a  game theoretic model with players in a large social network. We consider observations from one single equilibrium of a large network game with asymmetric information, in which each player chooses an action from a finite set and is subject to interactions with her friends.  Simple assumptions about the structure are made to establish the existence and uniqueness of equilibrium. In particular, we show that the equilibrium strategies satisfy a network decaying dependence (NDD) condition  requiring that dependence between any two players' decisions decays with their network distance.  The formulation of such an NDD property is novel and serves as the basis for statistical inference. Further, we establish the identification of the structural model and introduce a computationally feasible and efficient estimation method. We illustrate the estimation
method with an actual application to college attendance, as well as in Monte Carlo experiments.
\\

\textbf{Keywords:} Local interaction, social networks, Bayesian Nash Equilibrium, network decaying dependence condition, approximated maximum likelihood estimation \\

\textbf{JEL}: C14, C35, C62 and C72}

\begin{document}
\thispagestyle{empty}

\title{Social Interactions in Large Networks: A Game Theoretic Approach$^*$}

\author{\href{mailto:h.xu@austin.utexas.edu}{Haiqing Xu}}\date{ \today}\thanks{Department of Economics, The University of Texas at Austin, BRB 3.160, Mailcode C3100, Austin, TX, 78712 \href{mailto:h.xu@austin.utexas.edu}{h.xu@austin.utexas.edu}} 
\thanks{$^*$A previous version of this paper was circulated under the title ``Social interactions: a game theoretic approach''. I gratefully acknowledge    Kalyan Chatterjee, Sung Jae Jun, Isabelle Perrigne, Joris Pinkse, Quang Vuong and Neil Wallace for their guidance and advice.  I also thank Jason Abrevaya, Victor Aguirregabiria,  Edward Coulson, Stephen Donald, Frank Erickson, Bryan Graham, Ed Green, Paul Grieco, Han Hong,  Hiroyuki Kasahara, Konrad Menzel, Margaret Slade, James Tybout,  Halbert White, Daniel Xu, and the seminar participants  at UBC, Toronto, UT Austin, Boston College, Rice, Ohio State University, Texas A\&M, the 2011 Cornell--PSU Macro Seminar, the 2011 North American Econometric Society Summer Meeting, and 2012 Texas Econometrics Camp for providing helpful comments. All remaining errors are mine.}

\begin{abstract}
 \abstraction
\end{abstract}

\vspace{5ex}

\maketitle

\vspace{5ex}

\clearpage

\section{Introduction}

Over the last decades, network effects  on social behaviors  has become important in social theory \citep[see e.g.][]{granovetter1985economic}. In particular, economics has been encouraged to broaden its scope to the analysis of social interactions while maintaining the rigor that is emblematic of economic analysis \citep{manski2000economic}. Recently, game theory has  played a central role in this regard and a leading example is the study of network formation by \cite{bala2000noncooperative}. In this paper, we propose a network game of incomplete information to study large--network--based social interactions. Simple assumptions about the game structure are made to ensure a unique equilibrium and the equilibrium satisfies a decaying network effects condition. We then establish identification and estimation of the structural primitives using data from a single large network.

The structure of our network game follows the ``preference interaction'' approach suggested by \cite{manski2000economic}. Specifically, a player's payoff from choosing an action over alternatives  depends on other players' simultaneous actions as well.\footnote{In particular, we assume each player's payoff relevant covariates (including friend relationship) are public information, but payoff shocks (i.e. the error terms) are private information of the player.} Instead of interacting with all players on the social network, we assume each agent is only affected by the choices of her direct best friends. We call it as ``local'' interactions, a notion that was first introduced by \cite{Seim2006} in the context of industrial organization. Such a specification is parsimonious, but rich enough to generate the interdependence of all agents' choices, which is shaped by the way the network gets connected. For example,  teenagers are inclined to be affected by their friends in terms of adolescent risky behaviors \citep[see e.g.][]{nakajima2007measuring}, but such local effects can spread  through the network. In particular, they are indirectly affected by the behaviors of their friends' friends, because their friends are interacting with. In equilibrium, all teenagers from the same network will affect each other directly or indirectly.

Our local interaction specification differs from the ``linear--in--mean'' approach widely used in the literature on social interactions \citep[see e.g.][]{manski1993identification}. The latter captures the notion that an individual's behavior depends on the average behavior of all other social members.  The local interaction approach is attractive in the study of large--network--based social interactions: First, our model allows one to study the counterfactuals and the policy effects from the change of network graph. In contrast, much of the theoretical literature on network interventions have long focused attention on  qualitative features like the stability, but not quantitative effects.  A second advantage is that an equilibrium in our local interaction model exhibits features that reflect how the large network connects players to each other.  Last but not least,  the strategic effects between any pair of friends in our model are not diluted by the large size of the network, which is a typical feature in linear--in--mean models. 

By restricting the interaction strength to be sufficiently mild, we establish the uniqueness of the equilibrium. Uniqueness of the equilibrium is always crucial and of interest to both theoretical and empirical sides in game theory. In the presence of multiple equilibria, it is quite difficult to characterize the whole set of equilibria in a large network game: When the network is not regular, we cannot use a Markov--type of equilibrium solution concept to simplify empirical analysis like in the dynamic model inference. Another more fundamental concern is the ``incompleteness'' of the econometric model due to the existence of multiple equilibria \citep[see][]{Tamer2003}. In the empirical game literature, uniqueness  can also be found in, e.g., \cite{brock2001discrete} and  \cite{Xu2009}.

While there are strategic interactions among friends in a large network, players' choices are mutually dependent on each other. Intuitively, one would expect such a dependence to decay with the network distance. Under primitive conditions on the strategic components, we show that the dependence of a player's equilibrium choice on her friends' choices decays (exponentially) with network distance, a so--called ``{\it network decaying dependence condition}'' (NDD condition) that more or less amounts to the restrictions for a  stationary solution in the autoregressive model.  Our NDD condition is related to  a number of dependence decay conditions used in the time series and spatial analysis \citep[see e.g.][]{jenish2009central}. When the data come from the equilibrium of a single large network game, all observations are dependent on each other due to strategic interactions. The NDD implies that any two players' decisions are closer to be independent if they are farther away from each other. The formulation of NDD is novel and serves as the basis for our statistical inference.


For estimation, a key challenge arises because it is costly to solve the equilibrium analytically or numerically.  We propose a new approach that approximates the equilibrium solution by solving $n$ (i.e. the number of players) Bayesian games of much smaller size, one for each player. Specifically, for player $i$, a Bayesian game is tailored from the original one by cutting off all those players whose network distances from $i$ are larger than $h$ ($h\in\mathbb N$). The set of players left on the sub--network, as well as their payoffs, action space, information structure and so on, defines a smaller--sized Bayesian game. We solve this subnetwork game and use the equilibrium solution of player $i$ to approximate her equilibrium strategy  in the original large network game. The tuning parameter $h$ is chosen carefully depending on the network size $n$, i.e., $h$ needs to increase with $n$ at an exponential rate such that  approximation errors are negligible for the limiting distribution of the estimator. By this approximation, we then define an  approximated  maximum likelihood estimator (AMLE), which is shown to be asymptotically equivalent to the infeasible MLE. We  use Monte Carlo experiments to illustrate  the proposed estimation method, which performs well.

It is worth pointing out that our asymptotic analysis is based on the number of players going to infinity, instead of the infinite repetition of the same game with a small fixed number of players. The latter asymptotic approach is used by most of the existing empirical game literature, e.g. \cite{BjornVuong1984}, \cite{BresnahanReiss1991a}, \cite{brock2001discrete} and \cite{Tamer2003}. Our asymptotic analysis applies to observations coming from one or a small number of large networks.  In a a seminal paper, \cite{menzel2013large} characterizes the asymptotic distribution of a large matching market.  The analysis is similar in spirit to our approach in terms of using the limiting distribution as the number of players goes to infinity to approximate the distribution of the equilibrium in a large game. An important difference is that in \cite{menzel2013large} the strategic effects that cause the endogeneity issue become negligible as the number of players increases to infinity, which is not the case in our asymptotic analysis.


We apply our methods to study  college attendance decisions of high school students by using  the data from the National Longitudinal Study of Adolescent Health (Add Health). The Add health data is a longitudinal survey containing a nationally representative sample of adolescents in the United States during the 1994--95 school year. A unique feature in Add Health data is  the availability of respondents' social network information, which is reconstructed by students' best friends nominations in the survey. Applying the proposed estimation procedure, we find statistically significant, positive peer effects, which has a similar scale to other empirical findings of peer effects on youth behaviors using the similar or earlier datasets. See e.g. \cite{calvo2009peer} and \cite{gaviria2001school} .

The rest of the paper is organized as follows. In Section 2, we describe the data and provide descriptive statistics. In Section 3, we introduce the network game model and establish the uniqueness of the BNE and the NDD condition. In Sections 4, we establish identification of the model. In Section 5, we propose an estimation procedure. To present basic ideas, we first show the proposed estimator is asymptotically equivalent to the MLE and derive its asymptotic properties. We then study its finite sample performance by using Monte Carlo experiments. Applying the proposed method, we also present the baseline coefficient estimates and compares them with alternative empirical results in the literature. Section 6 concludes.  Proofs are provided in the Appendix.

\section{Data}

We study peer effects on college attendance of high school students using  data from the National Longitudinal Study of Adolescent Health (Add Health), which  is a longitudinal  survey containing a nationally representative sample of adolescents in the United States during the 1994--95 school year. A unique feature of the \textit{Add Health} data is the availability of respondents' social network information, as well as their social and economic characteristics (including college attendance):  Each respondent provides his or her friendship information by nominating at most five male and female best friends, respectively.  Intuitively, one can then reconstruct the whole friendship network among respondents.  All the respondents in our empirical study come from three high school networks and the total number of observations is $n=831$. A detailed description of the data can be found on the website of the Carolina Population Center.\footnote{See \url{http://www.cpc.unc.edu/projects/addhealth/data}.}

The college attendance  decisions must have been made by individual families during a short period. Following the literature \citep[see e.g.][]{christensen1975factors,leslie1988economic}, the exogenous covariates that affect college attendance include age, household income, GPA, parents' education level, race, gender, etc.  Descriptive statistics are presented in  \Cref{table5}.  The demographic variables, i.e., Household Income, Mother's Education and Father's Education, are recorded by some codes. These codes are natural numbers increasing with the actual value of variables. The median  Household Income is between \$50,000 and \$74,999. Mother's Education and Father's Education are coded as 1 = never went to school, 2 = not graduate from high school, 3 = high school graduate, 4 = graduated from college or a university, 5 = professional training beyond a four-year college. There is a severe missing data issue in these two variables: we treat missing observations as value 0. For the observed subsample, the medians of Mother's Education and the median of Father's Education are high school graduate. Over the whole sample, however, both medians are zero. 

As a matter of fact, we only use observations from three largest schools. For schools with small numbers of respondents, the missing data issue is severe. Therefore, the descriptive statistics in  \Cref{table5} are slightly different from other studies on social interactions that  use  the whole \textit{Add Health} dataset \citep[see e.g.][]{calvo2009peer}. 

The number of friends and the network centrality are two descriptive statistics on the network structure. Player $i$'s network centrality is defined by the number of players who take $i$ as a friend, i.e., $\sum_{j\neq i} \mathbb 1 (i \in F_j)$.  In the data, the standard deviation of the number of friends is less than the standard deviation of the network centrality, which is a typical feature in social networks. 
\begin{table}[h]
\small
\caption{Descriptive Statistics: 3 school networks; Year 1994--1995}
\label{table5}
\begin{center}
\begin{tabular}{lcccc}
\hline\hline
Variable& Mean& Std. Dev.&Min&Max\\\hline
Age&17.088&1.138 & 15&21\\
Female&0.502&0.500&0&1\\
Household Income&8.827&2.122& 1 & 12\\
Mother's Education*&0.516&1.676&0&11\\
Father's Education&1.709&2.955&0&12\\
Overall GPA&2.376&0.772& 0.11&4\\
American Indian**&0.039&0.193&0&1\\
Asian&0.140&0.347&0&1\\
Black&0.084&0.278&0&1\\
Hispanic&0.348&0.477&0&1\\
White&0.651&0.477&0&1\\
Other Race& 0.153 & 0.360 &0&1 \\
Number of Friends & 1.303&1.575 & 0 & 8\\
Network Centrality & 1.303 &1.780&0 & 13 \\\hline
College Attendance &0.535&0.499&0&1\\
\hline
\multicolumn{5}{l}{\footnotesize{*Missing observations have been treated as 0.}}\\
\multicolumn{5}{l}{\footnotesize{**Note that some observations are associated with more than one race.}}
\end{tabular}
\end{center}
\end{table}

 \section{The Model}
 \label{themodel}
Following our empirical application, we consider a game theoretic model on social interactions of high school students' college attendance decisions.  All these students are denoted as players indexed by $i\in  N\equiv \{1,\cdots,n\}$, with exogenously determined locations on the school network. Using the terminology in graph theory, a vertex of the network  denotes a student and a directed edge connects vertex $i$ to $j$ if student $j$ is one of $i$'s best friends.  Following the network distribution theory \citep[see e.g.][]{barabasi1999emergence}, we can view the high school network as a random graph with vertex connectivities governed by some probability distribution, and the observed network in our data is a single realization of the large random network. Given the network, we denote $F_i$ as the group of  $i$'s best friends, i.e., the set of students are directly connected to $i$.    Note that friendship may not be symmetric, i.e., $j\in  F_i$ does not necessarily imply $i\in F_j$, which is an important feature in our data. Moreover, let  $Q_i=\#  F_i$ be the number of $i$'s best friends.   In our game theoretic model,  we assume the school network structure is public information. Therefore, $F_i$ is also public information. 

Moreover, we assume each player $ i$ simultaneously chooses a discrete action $Y_i\in {A}\equiv \{0,1,2,...,K\}$. Following the convention, let $Y_{-i}$ denote a profile of actions of all other players. Let further  $X_i\in\mathscr S_X\subseteq \mathbb{R}^{d}$ be a vector of player $i$'s payoff relevant state variables, which are publicly observed by all players, as well as the researcher.  Further, player $i$ observes a vector of action--specific payoff shocks labeled as $\epsilon_i\equiv (\epsilon_{i0},\cdots,\epsilon_{iK})\in\mathbb R^{K+1}$. We assume that $\epsilon_i$ is $i$'s private information, i.e., $\epsilon_i$ is not observed by any $j\neq i$.\footnote{It should be noted our specification rules out unobserved
heterogeneity, which is observed by all the players but not the researcher.}   In our empirical application, $Y_i$ is  binary indicating colleague attendance, where action $1$ denotes the college attendance; $X_i$ is a vector of demographic variables including e.g. age, gender, GPA, parents' education, household income and race;  Moreover, $\epsilon_i$ is an idiosyncratic preference shock for college attendance. For expositional simplicity, we denote all the public state variables associated with student $i$ by $S_i\equiv (X_i',F_i)'$.

Players interact with each other through their utilities. Specifically, we specify player $i$'s payoff from choosing an action $k\in A$ as follows
 \begin{equation}
 \label{payoff1}
 U_{ik}( Y_{-i},S_i, \epsilon_i)=\beta_k(X_i)+\sum_{j\in  F_ i}\alpha_k( Y_j,X_i, Q_i)+\epsilon_{ik}, 
 \end{equation}
 where  
 $\beta_k(\cdot)$ is a choice--specific function, and $\alpha_k(\cdot,\cdot,\cdot)$ measures the strategic effects on  $i$'s payoff (of choosing $k$) from  her friend $j$'s decision. In our specification, the strategic effects depend on the state variable $X_i$ as well as  $Q_i$. Because only the differences of choice--specific payoffs matter to decision makers, hence, w.l.o.g., we normalize the mean utility of action $0$ by letting $\beta_0(x)=\alpha_0(\ell, x, q)=0$ for all  $x\in\mathscr S_X$, $\ell\in{A}$ and $q\in\mathbb N$. Let $\theta_k=(\beta_k,\alpha_k)'$ and $\theta=(\theta'_1,\cdots,\theta'_K)'$ be the structural parameters of the game, which are unknown functions. 
 
It is worth pointing out that our model can be extended to allow for exogenous interaction effects, i.e., player $i$'s payoffs $U_{ik}$ depends on $X_j$ for all $j\in F_i$. See e.g. \cite{manski1993identification} and \cite{bramoulle2009identification}.  Our approach could be modified to accommodate such an extension.\footnote{We thank a referee for this point.}  In our empirical application on college attendance, high school students are less likely to make their attendance decisions according to friends' demographic variables (e.g. Household income, Parents' education level and Overall GPA). On the other hand, such payoff relevant covariates of friends can  affect  each individual's decision indirectly through her expectation on friends' equilibrium choices. 



 In our setting, direct interactions on payoffs only occur among friends. Although interactions are local, strategic effects can spread throughout the whole network if no subnetwork is isolated. For instance, each player needs to consider the decisions by the friends of her friends. This is because those decisions are relevant to her friends' choices which thereafter affect her payoffs. In the equilibrium, each player's strategy depends on all other players' public observables $\{(X_j,F_j)\}_{j\neq i}$ as well as her own state  variables $(X_i,F_i)$.\footnote{A recent work by  \cite{manresa2013estimating} develops a reduced form to assess the  dependence structure from social interactions in a linear setting.}

\subsection{Bayesian Nash Equilibrium (BNE)}

Let $\mathbb S_n=(S_1,\cdots,S_n)$ be all the public information in the network game. For simplicity, we will suppress the subscript $n$ in $\mathbb S_n$ unless the subscript is necessary. To discuss the equilibrium solution, we fix  the public state variable  $\mathbb S$.

In this Bayesian game,  player $i$'s strategy  is a function $r_i(\cdot|\mathbb S; \theta)$ that maps her private information $\epsilon_i$ to a discrete choice $Y_i$. Following the BNE solution concept,  player $i$'s equilibrium strategy, denoted as $r^*_i$, maximizes her (conditional) expected payoff given all other players equilibrium strategies $r^*_{-i}$, i.e.,
 \begin{multline}
 \label{payoff}
r^*_i(\epsilon_i|\mathbb S; \theta)=\argmax_{k\in{A}} \mathbb{E}\left[U_{ik}(Y_{-i},S_i,\epsilon_i)|\mathbb S,\epsilon_i\right]\\
=\argmax_{k\in{A}} \left[\beta_k(X_i)+\sum_{\ell=0}^K\Big\{\alpha_k(\ell,X_i, Q_i)\times\sum_{j\in F_i}\mathbb P\left(r_j^*(\epsilon_j|\mathbb S; \theta)=\ell\big|\mathbb S,\epsilon_i\right)\Big\}+\epsilon_{ik}\right], \ \ \forall i.
 \end{multline} 
 Thus, \cref{payoff} defines a simultaneous equation system in terms of $\big(r_1^*,\cdots,r_n^*\big)$. 
 
To characterize the BNE solution, we first make the following assumption on the private information $\epsilon_i$.
 \begin{assumption}
\label{extremevalue}
Let $\epsilon_{ik}$ be i.i.d.  across both actions and players and conform to an extreme value distribution with a density function 
$
f(t)=\exp\left(-t\right)\exp\left[-\exp\left(-t\right)\right].
$
\end{assumption}
 \noindent
\Cref{extremevalue} has been widely assumed in the discrete choice model literature, as well as in empirical discrete games \cite[see, e.g.,][]{brock2002multinomial,bajari2010estimating}. The independence of $\epsilon_i$ across players in \cref{extremevalue} implies that players' equilibrium choices are conditionally independent given $\mathbb S$. Therefore, the network dependences of players' decisions are all characterized by the dependence of players equilibrium strategies  $r_i^*$ on the common state variable $\mathbb S$. 

By \cref{extremevalue}, we can rewrite (\ref{payoff})  in terms of equilibrium choice probabilities. Let $\sigma^*_{ik}(\mathbb S; \theta)=\mathbb P\left(r^*_i(\epsilon_i|\mathbb S; \theta)=k|\mathbb S\right)$  and $\sigma^*_i(\mathbb S; \theta)=\left(\sigma^*_{i0}(\mathbb S; \theta),\cdots,\sigma^*_{iK}(\mathbb S; \theta)\right)'$ be the equilibrium choice probabilities of action $k$ and the action profile, respectively. Let further  $\Sigma^*(\mathbb S; \theta)=(\sigma^*_1(\mathbb S;\theta),\cdots,\sigma^*_n(\mathbb S;\theta))$ be the equilibrium choice probability profiles of all players.   By (\ref{payoff}) and \cref{extremevalue}, we have 
\begin{equation}
 \label{choiceprobability}
\sigma^*_{ik}(\mathbb S; \theta)=\frac{\exp\left[\beta_k(X_i)+\sum_{l=0}^K\left\{\alpha_k(\ell, X_i,Q_i)\times\sum_{j\in  F_i}\sigma^*_{j\ell}(\mathbb S; \theta)\right\}\right]}{1+\sum_{q=1}^K\exp\left[\beta_q(X_i)+\sum_{l=0}^K\left\{\alpha_q(\ell, X_i,Q_i)\times\sum_{j\in  F_i}\sigma^*_{j\ell}(\mathbb S; \theta)\right\}\right]}, \ \forall i, k.
\end{equation}
Note that solving the BNE solution  $\{r^*_1(\cdot|\mathbb S; \theta), \cdots, r^*_n(\cdot|\mathbb S; \theta)\}$ to  \cref{payoff} is equivalent to solving $\{\sigma^*_1(\mathbb S; \theta), \cdots, \sigma^*_n(\mathbb S; \theta)\}$ from (\ref{choiceprobability}). See \cite{bajari2010estimating}. 

 \Cref{choiceprobability} is the common logit functional form, except for the presence of the equilibrium choice probabilities of $i$'s friends.   The existence of a solution follows Brouwer's fixed point theorem. Next, we establish the uniqueness of the equilibrium, and then show the equilibrium satisfies  a  decaying dependence condition. Both uniqueness and the decaying dependence condition are crucial for our empirical analysis. 


\subsection{Unique equilibrium}

The insight for deriving the unique equilibrium comes from the linear spatial autoregressive model literature: Strong interactions among individuals  can induce multiple equilibria in a simultaneous equation system.  To obtain the uniqueness of the BNE, we need to restrict the interaction strength to be sufficiently mild.

 \begin{assumption}
\label{smallstrategyeffect}
Denote $\lambda\equiv \frac{K}{K+1}\cdot \sup_{(x,q)\in \mathscr S_{XQ}}\max_{\ k,m, \ell\in A}\ q|\alpha_k(\ell, x,q)-\alpha_m(\ell,x,q)|$. Let $\lambda<1$. 
\end{assumption}
\noindent
For estimation,  we parametrize $\alpha_k(\ell,x,q)$ by $\alpha_{k\ell}/q$ for some $\alpha_{k\ell}\in\mathbb R$.  Then, \cref{smallstrategyeffect} becomes 
\[
\max_{k,m, \ell\in A}\  |\alpha_{k\ell}-\alpha_{m\ell}|<(K+1)/{K}.
\]Similar to the requirement that all roots lie outside of the unit circle  in spatial autoregressive models,  such a condition ensures weak dependence. 

In our empirical application,  each student takes a binary decision for college attendance. Under the parametrization,  \cref{smallstrategyeffect} can be rewritten as 
\[
\max\  \{|\alpha_{10}-\alpha_{00}|, |\alpha_{11}-\alpha_{01}|\}< 2,
\]
Note that $\alpha_{00}-\alpha_{10}$ and $\alpha_{11}-\alpha_{01}$ describe peer effects in social interactions, i.e., the principal that friends benefit from choosing the same action. Intuitively,  $\alpha_{00}-\alpha_{10}\geq 0$ and $\alpha_{11}-\alpha_{01}\geq 0$ in our empirical context. Therefore, \cref{smallstrategyeffect} requires peer effects to be bounded above. Intuitively, this condition means that the college attendance decisions are mainly determined by students' own social and economic characteristics  like GPA, household income etc., and their idiosyncratic preference shock on college attendance as well. If the average probabilities of friends' college attendance increase one percentage point, then the peer effects on her own college attendance probability is limited by $\lambda<1$ percentage.\footnote{To see this, note that \cref{smallstrategyeffect} ensures a quasi--Lipschitz condition hold for the best response function:  The best response function $\Gamma_i(s_i, \{\sigma_j: j\in F_i\})$ defined by \eqref{gamma} in the appendix satisfies the following condition:
\[
\|\Gamma_i(s_i, \{\sigma_j: j\in F_i\})-\Gamma_i(s_i, \{\tilde \sigma_j: j\in F_i\})\|_1\leq \lambda \cdot \max_{j\in F_i} \|\sigma_j-\tilde \sigma_j\|_1
\]where $\|\cdot \|$ is the $L^1$--norm.  See the proof of \Cref{uniqueness}. } 

\Cref{smallstrategyeffect} generally holds in a wide range of empirical studies of youth behaviors, including e.g. the substance use, church attendance, academic performance, academic cheating. See e.g., \cite{gaviria2001school}, \cite{sacerdote2001peer},  \cite{kawaguchi2004peer}, \cite{carrell2008peer} and  \cite{calvo2009peer}.  In these studies, the  effects on a player's equilibrium choice probabilities from her friends' choices are significantly smaller  than one.  Although the NDD is a natural condition for peer effects in our empirical context, numerous prominent exceptions exist. For example, adolescent risky behaviors like substance (marijuana, alcohol, or tobacco) use are mainly driven by influence from friends. See e.g. \cite{gaviria2001school} and \cite{kawaguchi2004peer}. 
Another leading example is the butterfly effects widely used in e.g. fashion, financial crisis and gold rush,  which characterize the sensitive dependence of players' choices on each other.

\begin{lemma}
\label{uniqueness}
Suppose  \cref{extremevalue,smallstrategyeffect} hold. Then,  there always exists a unique BNE, regardless of the number of players $n$ or the realization of the state variable $\mathbb S$. 
\end{lemma}
The proof of the uniqueness of the BNE relies on a contraction mapping argument. We can  generalize such a result to the exponential family distribution for the private information $\epsilon_i$.  

\subsection{Network Decaying Dependence (NDD)}
We begin with some notation. For  any positive integer $h\in\mathbb N$, let $N_{(i,h)}$ be the subset of players defined inductively:
\[
N_{(i,0)}= \{i\}\ \ \ \text{and}\ \ \forall \ h\geq 1, \ N_{(i,h)}=N_{(i,h-1)}\bigcup\Big(\bigcup_{j\in N_{(i,h-1)}}  F_{j}\Big).
\]By definition,  $N_{(i, h)}$ is the set of players on the social network within $h$ distance of player $i$ (including $i$ herself).  Moreover, let $\mathbb G_{(i,h)}$ be the network graph that uses vertices and edges to describe all the connections among $N_{(i, h)}$. Let further $\mathbb S_{(i,h)}=\left(\{X_j: j\in N_{(i,h)}\};\mathbb G_{(i,h)}\right)$. By definition, $\mathbb S_{(i,h)}$ describes the subnetwork centered around player $i$ within her $h$--distance, i.e., how these players are connected to each other and what are the state variables at each node of the graph.  Note that players' identities do not matter in the definition of $\mathbb S_{(i,h)}$. 


The idea of NDD condition is to examine how player $i$'s equilibrium choice probability $\sigma^*_i(\mathbb S ; \theta)$ responds to  counterfactual changes of some other player $j$'s public state variable $S_j$.  Note that in equilibrium $\sigma^*_i(\mathbb S ; \theta)$ depends on all the public information $\mathbb S$, including  $S_j$ no matter $j$ is $i$'s friend or not. In a ``stable'' equilibrium,  intuitively such a dependence should decay with distance. Therefore, the statistical dependence between $Y_i$ and $Y_j$ also diminishes with $i$ and $j$'s network distance. 

\begin{definition}[Network Decaying Dependence, NDD]
\label{def_1} In the above network game, the equilibrium satisfies the NDD condition if there exists a deterministic sequence $\{\xi_h:h=1,\cdots,\infty\}$ with  $\xi_h\downarrow 0$ as $h\rightarrow \infty$ such that  for any given size $n$ of the network and positive integer $h$, we have
\begin{equation}
\label{stable}
\sup_{s,s'\in\mathscr S_{\mathbb S}: \ s_{(i,h)}=s'_{(i,h)}}\left\|\sigma^*_i(s; \theta)-\sigma^*_i\left(s'; \theta\right)\right\|_1\leq \xi_{h},\ \ \forall i=1,\cdots, n,
\end{equation}where $\|\cdot\|_1$ is the $L_1$--norm, i.e., for any $z\in\mathbb R^k$, $\|z\|_1=\sum_{\ell =1}^k |z_\ell|$.
\end{definition}
\noindent
Our notion of NDD is related to the weak dependence in the time--series/spatial literature. In particular, NDD implies the near--epoch dependence (NED) condition in e.g.  \cite{andrews1988laws}.\footnote{This is because by \cref{stable},  $\mathbb P\big(Y_i\neq \mathbb E [Y_i|\{(\epsilon_j, S_j): j\in N_{(i,h)}\}]\big)$ is bounded by $K\xi_{h} $.} Different from the  time-series/spatial statistical literature that assumes weak dependence of unobserved errors across observations,   the dependence of players' decisions  results from network--based strategic interactions. Conditional on $\mathbb S$, players' decisions are mutually independent under  \cref{extremevalue}. 

In \Cref{def_1}, NDD requires the causal effects of $S_j$ on $\sigma^*_{i}$ to be bounded above by $\xi_{\rho(i,j)}$, where $\rho(i,j)$ denotes the network distance from $j$ to $i$. The game size $n$ is treated as a state variable. With NDD (and \cref{maximumneighbours} to be introduced later), if we increase the network size by keeping adding players to the ``fringe'' of the network, then the equilibrium choice probability for any existing player will converge to a limit.\footnote{To see this, fix $i$ ($i\leq n$) and consider adding new players $n+1,n+2,\cdots$ one by one to an existing network with $n$  players. \Cref{maximumneighbours} ensures that  for any existing player $i\leq n$, the network distance from the added player $n+k$ to $i$, i.e., $\rho(i,n+k)$, will go to infinity as $k$ goes  to infinity, if any new player is added to the fringe of the previously existing network (i.e., a new player will not decrease the network distance of any pair of existing players). Therefore, $\{\sigma^*_i(\mathbb S_{ n'};\theta): n'=n, n+1,\cdots,\infty\}$ is a Cauchy sequence if the NDD condition holds. } The next lemma shows that the equilibrium in our network game satisfies NDD under weak conditions. 
\begin{lemma}
\label{stability}
Suppose  \cref{extremevalue,smallstrategyeffect} hold.  Then the BNE satisfies NDD with $\xi_h=2\lambda^{h+1}$. 
\end{lemma}
\noindent
With the NDD,  the distribution of the observable variables  $\mathbb P_{Y_i|\mathbb S}$ can be nonparametrically estimated by using data from one single large network as the network size $n$ goes to infinity. See \Cref{appendixd}.

\section{Identification}\label{section3}
In this section, we  discuss the identification of the structural parameter $\theta$. Following \cite{hurwicz1950generalization} and \cite{koopmans1950identification}, the definition of identification in a structural model requires that there is a unique value of the structural parameter $\theta$ that generates the distribution of the observable variables  $\{\mathbb P_{Y_i|\mathbb S}: i=1,\cdots,n\}$.

Because of the uniqueness of the equilibrium by \Cref{uniqueness}, $\sigma^*_{ik}(\mathbb S; \theta)$ is identified by $\sigma^*_{ik}(\mathbb S; \theta)=\mathbb P(Y_i=k|\mathbb S)$.  Let $\delta_{ik}(\mathbb S)=\ln\mathbb P(Y_i=k|\mathbb S)-\ln\mathbb P(Y_i=0|\mathbb S)$ for each $k\in\mathcal{A}$.  By definition, $\delta_{ik}(\mathbb S)$ is also identified. Moreover, by (\ref{choiceprobability}),  
\begin{equation*}
\delta_{ik}(\mathbb S)=\beta_k(X_i)+\sum_{\ell\in A}\Big[\alpha_k(\ell,X_i,Q_i)\times\sum_{j\in F_i}\mathbb P(Y_j=\ell|\mathbb S)\Big], \ \forall i, k.
\end{equation*}
 Let  further $\phi_{i\ell}(\mathbb S)=\sum_{j\in F_i}\mathbb P(Y_j=\ell|\mathbb S)$. By definition, $\sum_{\ell\in A}\phi_{i\ell}(\mathbb S)=Q_i$. It follows that
 \begin{equation}
 \label{logchoiceprobability}
 \delta_{ik}(\mathbb S)=\beta_k(X_i)+\alpha_k(0,X_i,Q_i)\times Q_i+\sum_{\ell=1}^K\big\{[\alpha_k(\ell,X_i,Q_i)-\alpha_k(0,X_i,Q_i)]\times\phi_{i\ell}(\mathbb S)\big\}.
 \end{equation}
Similar to  \cite{robinson1988root}, \cref{logchoiceprobability} is essentially a partial linear model as shown in \Cref{pointidentification}. 

Equation (\ref{logchoiceprobability}) suggests that $\beta_k(\cdot)$ and $\alpha_k(0,\cdot,\cdot)$ are not identified separately unless $0\in\mathscr S_Q$.\footnote{To see this, consider the following  specification: $\alpha_k(0,X_i,Q_i)=\alpha_k(0,X_i)/Q_i$ for all $Q_i\geq 1$.} Hence, we introduce the following normalization on $\alpha_k$. 
 \begin{assumption}
 \label{payoff_normalization}
Let $\alpha_k(0,\cdot,\cdot)=0$ for all $k\in A$. 
 \end{assumption}

Next, we assume a rank condition for identification.  Let   $\varphi_i(\mathbb S)=\left(1, \phi_{i1}(\mathbb S),\cdots,\phi_{iK}(\mathbb S)\right)'$.
\begin{assumption}[Rank Condition]
\label{identification}
Given the game size $n$, the matrix $\mathbb E[ \varphi_i(\mathbb S)\times \varphi_i(\mathbb S)'|X_i=x,Q_i=q]$ is invertible for all $(x,q)\in\mathscr S_{XQ}$.
\end{assumption}
\noindent
\Cref{identification} is  testable given that the conditional choice probabilities
can be consistently estimated.
 
 
The next theorem establishes the identification of the model. For the sake of simplicity, let $\alpha_k(\cdot,\cdot)=(\alpha_k(1,\cdot,\cdot),\cdots,\alpha_k(K,\cdot,\cdot))'$ be a vector of functions. 
\begin{lemma}
\label{pointidentification}
Fix  arbitrary $n$. Suppose \cref{extremevalue,smallstrategyeffect,payoff_normalization,identification} hold. Then the structural parameter $\theta$ is identified, i.e. $\mathbb P_{Y_1,\cdots,Y_n|\mathbb S}(\theta')\neq\mathbb P_{Y_1,\cdots,Y_n|\mathbb S}(\theta)$ for all $\theta'\neq\theta$. Specifically, for any $(x,q)\in\mathscr S_{XQ}$,
\[
 \Bigg(\begin{array}{c}\beta_k(x) \\ \alpha_k(x,q)  \end{array}\Bigg)=\left\{\mathbb{E}\left[\varphi_i(\mathbb S)\times\varphi'_i(\mathbb S)|X_i=x,Q_i=q\right]\right\}^{-1}\mathbb{E}\left[\varphi_i(\mathbb S) \times\delta_{ik}(\mathbb S)|X_i=x,Q_i=q\right].
\]
\end{lemma}

\noindent
Note that the identification result in \Cref{pointidentification} is established for each fixed $n$. For purpose of estimation and asymptotic analysis, however we need $n$ goes to infinity. Hence, we  replace the rank condition \ref{identification} by the following assumption.
\begin{assumption}[Rank Condition for large $n$] The matrix $\mathbb E\left[ \varphi_i(\mathbb S)\times \varphi_i(\mathbb S)'|X_i=x,Q_i=q\right]$ is invertible for all $n$ sufficiently large and $(x,q)\in\mathscr S_{XQ}$, i.e., for any $(x,q)\in\mathscr S_{X_i,Q_i}$,
\label{idlimit1} 
\[
\liminf_{n\rightarrow\infty} \ \Det\Big(\mathbb E\left[ \varphi_i(\mathbb S)\times \varphi_i(\mathbb S)'|X_i=x,Q_i=q\right]\Big)>0.
\] 
\end{assumption}

By relaxing conditions in \Cref{pointidentification},  the next theorem establishes identification of the model for all sufficiently large $n$.  
\begin{theorem}
\label{idinlimit}
Suppose \cref{extremevalue,smallstrategyeffect,payoff_normalization,idlimit1} hold. Then the structural parameter $\theta$ is identified for all $n$ sufficiently large. 
\end{theorem}

The semiparametric identification in \Cref{idinlimit} helps the applied researcher to get a better sense of whether a fully parametric approach relies on  ad hoc specification (of the payoff function) for identification or merely for simplicity of estimation. Analogous rank condition can be formulated in the fully parametric model that is used for our estimation.  Let $\beta_k(x)=x'\beta_{k}$  and $\alpha_k(\ell,x,q)=\alpha_{k\ell}/q$, where  $\beta_{k}\in\mathbb{R}^{d}$ and $\alpha_k\equiv (\alpha_{k1},\cdots,\alpha_{kK})'\in\mathbb R^K$. Let $W_{i}=\left( X'_i, \phi_{i1}(\mathbb S),\cdots,\phi_{iK}(\mathbb S)\right)'$. 
\begin{assumption}[Rank Condition for linear--index setup]  
\label{para-id}
The matrix $\mathbb{E}\left(W_iW_i'\right)$ is invertible for all $n$ sufficiently large.
 \end{assumption}
\noindent 
Replace  \cref{idlimit1} with \ref{para-id} in \Cref{idinlimit}, then the identification of $\theta_k=(\beta'_{k},\alpha'_k)'$ is given as follows: for sufficiently large $n$, 
\[
\theta_k= \left[\mathbb{E}(W_iW_i')\right]^{-1}\mathbb{E}\left[W_i\delta_{ik}(\mathbb S)\right].
\] Clearly, variations in the aggregated friends' choice probabilities $\phi_{i\ell}(\mathbb S)$ identify the strategic coefficients $\alpha_{k\ell}$. 

\section{Estimation}\label{estimation}

This section discusses the parametric estimation of the structural parameter $\theta$.  In particular, we specify the payoff function by
 \begin{equation}
 \label{payoff2}
 U_{ik}( Y_{-i},S_i, \epsilon_i)=X_i' \beta_{k}+ \sum_{\ell=1}^K\alpha_{k\ell}\times\frac{1}{Q_i}\sum_{j\in  F_ i}\one (Y_j=\ell)+\epsilon_{ik} 
 \end{equation}
where $\beta_{k}\in\mathbb R^{d}$ and $\alpha_k=(\alpha_{k1},\cdots,\alpha_{kK})'\in\mathbb R^K$. Let $\theta_k=(\beta'_k,\alpha'_k)\in\mathbb R^{K+d}$. Moreover, let  $\beta=(\beta'_{1},\cdots,\beta_K')'\in\Theta_\beta\subseteq \mathbb R^{Kd}$ and $\alpha=(\alpha_1',\cdots,\alpha_K')'\in\Theta_\alpha\subseteq \mathbb R^{K\times K}$, where $\Theta_\beta$ and $\Theta_\alpha$ are the parameter space for $\beta$ and $\alpha$, respectively. Denote   $\theta=(\theta_1',\cdots,\theta_K')'$ and $\Theta= \Theta_\beta\times \Theta_\alpha $.

Let $\{X_i,F_i,Y_i\}_{i=1}^n$ be the data generated from the equilibrium of a single large network.  For asymptotic analysis, we consider the network size $n$ goes to infinity, since our empirical application  involves a few large networks. For the data generating process, our asymptotics requires the probability distributions of $\mathbb G_{(i,h)}$ with arbitrarily fixed $h$ converges to the same limiting distribution for all $i$ as $n\rightarrow\infty$, and the random graph $\mathbb G_{(i,h)}$ is independent of  $\mathbb G_{(j,h)}$ given they don't contain any common element. 

We now proceed to  motivate our estimation procedure. First, note that under \Cref{extremevalue}, the actions chosen by the players are conditional i.i.d. given $\mathbb S$. Thus, we have the (conditional) loglikelihood function
\begin{equation}
\label{likelihood}
\hat L(\theta)=\frac{1}{n}\sum_{i=1}^n\sum_{k\in A}\one(Y_i=k)\cdot \ln \sigma^*_{ik}(\mathbb S; \theta).
\end{equation} Let $\hat\theta_{MLE}=\argmax_{c\in\Theta} \ \hat L(c)$ be the MLE, which requires to solve $\{\sigma^*_{ik}(\mathbb S;\theta): i\in N; k\in A\}$ to \eqref{choiceprobability}.  In \eqref{likelihood},  we can verify that all the regularity conditions hold under additional weak conditions.\footnote{For instance, \Cref{c5} in the appendix ensures the differentiability of the objective function.}  In practice,  however, $\hat \theta_{MLE}$ is not computationally feasible when the network is large. This is because the equilibrium choice probability $\sigma^*_{ik}(\mathbb S; \theta)$ has no closed--form expression and its numerical solution is costly to obtain in the large simultaneous equation system. 

The key to our approach is to approximate $\sigma^*_{ik}(\mathbb S; \theta)$ by some computable solution $\sigma^{h}_{ik}(\mathbb S; \theta)$ to be defined later, where  $h$ is an integer that depends on $n$ such that the approximation error  $\|\sigma^{h}_{ik}(\mathbb S;\theta)-\sigma^*_{ik}(\mathbb S; \theta)\|_1$ is negligible relative to the sampling error. Thus, we define our approximated loglikelihood function
\begin{equation}
\label{aobj}
 \hat Q(\theta)=\frac{1}{n}\sum_{i=1}^n\sum_{k\in A}\one(Y_i=k)\cdot \ln \sigma^{h}_{ik}(\mathbb S\ ;\theta).
\end{equation}
Further,  our estimator maximizes the approximated likelihood, i.e., $\hat\theta=\argmax_{c\in\Theta} \ \hat Q(c)$.

To define $\sigma^h_{ik}(\mathbb S; \theta)$,  we first define a Bayesian game of smaller size: let $N_{(i,h)}$ be the set of players and each player $j\in N_{(i,h)}$ simultaneously makes a discrete choice $Y_j\in A$. Moreover, each player $j$ in $N_{(i,h)}$ has the same state variables  $(X_j,\epsilon_j)$ as the original network game, but player $j$'s set of friends is restricted to be $F_j\cap N_{(i,h)}$. In other words, we artificially removes all the players outside of $N_{(i,h)}$ in the original game. Note that player $i$ is located at the center of the subnetwork $N_{(i,h)}$.  Similarly, $\{\sigma^h_{ik}(\mathbb S; \theta): j\in N_{(i,h)}, k\in A\}$ solves:  
\begin{equation}
\label{eq_smallgame}
\sigma_{jk}=\frac{\exp\left[\beta'_{k}X_j+\sum_{\ell=1}^K\alpha_{k\ell}\times \left(\frac{1}{Q_j}\sum_{j'\in F_j\cap N_{(i,h)}}\sigma_{j'\ell}\right)\right]}{1+\sum_{q=1}^K\exp\left[\beta'_{q}X_j+\sum_{\ell=1}^K\alpha_{q\ell}\times\left(\frac{1}{Q_j}\sum_{j'\in  F_j\cap N_{(i,h)}}\sigma_{j'\ell}\right)\right]}.
 \end{equation}By \Cref{uniqueness}, there is a unique solution to (\ref{eq_smallgame}).  In this derived subnetwork game, player $i$ is in the center of the subnetwork and her  equilibrium choice probabilities  profile is denoted by $\sigma^{h}_{i}(\mathbb S; \theta)=\big(\sigma^{h}_{i0}(\mathbb S; \theta), \cdots, \sigma^{h}_{iK}(\mathbb S; \theta)\big)$. 
 
By \Cref{stability}, the approximation error  $\|\sigma^*_{i}(\mathbb S;\theta)-\sigma^h_{i}(\mathbb S;\theta)\|_1$ can be bounded by $2\lambda^{h+1}$.\footnote{To apply Lemma 2, let $\mathbb S_{i,h}$ denote the state  of the network derived from $\mathbb S$ by eliminating all the network connections outside of $N_{(i,h)}$. By definition, $\mathbb S_{i,h}\in\{s'\in\mathscr S_S: s'_{(i,h)}=\mathbb S_{(i,h)}\}$. Moreover, note that $\sigma^h_{i}(\mathbb S;\theta)=\sigma^*_{i}(\mathbb S_{i,h}; \theta)$, since all players outside of $N_{(i,h)}$ have no strategic effects on players in $N_{(i,h)}$.   } 
To control for the approximation error, we choose $h$ to increase with $n$ at a proper rate.\footnote{Note that for  $h=0$, the proposed estimator becomes the classical multinomial logit estimator. }

\subsection{Asymptotic analysis}\label{mlepara}
We now establish the consistency and limiting distribution for the proposed estimator. First, we make the following assumptions.  
\begin{assumption}
\label{compact}
(i) The parameter space $\Theta$ is compact and the support $\mathscr S_{XQ}$ is bounded; (ii) The true parameter $\theta$ belongs to the interior of $\Theta$.
\end{assumption}
\begin{assumption}
\label{stable2}
Let
\[
\sup_{a\in \Theta_\alpha} \max_{k,\ell,m\in A}|a_{k\ell}-a_{m\ell}|< {(K+1)}/{K}.
\]
\end{assumption}
\begin{assumption}
\label{SNdistr}
Given any $h\in\mathbb N$,  the probability distribution of $\mathbb G_{(i,h)}$ converges to a limiting distribution $\mathbb P_{\mathbb G,h}$ as $n\rightarrow \infty$ for all $i$; and $\mathbb G_{(i,h)}$ is independent of  $\mathbb G_{(j,h)}$ if $ N_{(i,h)}\cap N_{(j,h)}=\emptyset$. Moreover, the payoff covariates $X_i$ are i.i.d. across players given the exogenous random network. 
\end{assumption}
\begin{assumption}
\label{maximumneighbours}
There exists a positive constant $c_0\in\mathbb N$, which does not depend on $n$, such that $\max_{i\in N} \sum_{j\neq i} \one(i\in F_j)\leq c_0$ with probability one.
\end{assumption}
\begin{assumption}
\label{choiceh}
(i )Let $h\rightarrow \infty$ as $n\rightarrow\infty$; (ii) Let further $h=[h_0\cdot n^{a}]$ for some constant $h_0>0$ and $a>0$, where $[t]$ is the largest integer that is no larger than $t$.
\end{assumption}
Let $P= K(d+K)$ denote the dimension of the parameter $\theta$.  Moreover, let $f_{i}(Y_i| \mathbb S; \theta)=\sum_{k\in A}\one(Y_i=k)\times \ln\sigma^*_{ik}(\mathbb S; \theta)$ and $J_n(\theta)=\mathbb{E}\big[{\frac{\partial}{\partial \theta} f_{1}(Y_1|\mathbb S ;\theta)}\times {\frac{\partial}{\partial \theta'} f_{1}(Y_1|\mathbb S ;\theta)}\big]$.  The latter is indexed by $n$ because of the dependence of  $f_i$ on $n$ through $\mathbb S$ and $\sigma^*_i$.
\begin{assumption}
\label{JN}
There exists a non--singular $P\times P$ matrix $J(\theta)$ such that $J_n(\theta)\rightarrow J(\theta)$.
\end{assumption}
\noindent
\Cref{compact}--(i) ensures that choice probabilities are bounded away from zero so that the loglikelihood function is bounded. Unbounded regressors can be accommodated using high order moments restrictions \citep[see e.g.][]{van1990estimating}. \Cref{compact}--(ii)  is standard in the literature.  \Cref{stable2} strengthens \cref{smallstrategyeffect} to hold for all the  values in the parameter space $\Theta$.

\Cref{SNdistr,maximumneighbours} impose restrictions on the distribution of the state variables as well as the network connections. For the first half of \Cref{SNdistr}, note that for any given $n$ and $h$, the probability distribution of $\mathbb G_{(i,h)}$ is well defined, since the subgraph $\mathbb G_{(i,h)}$ is a mapping to the space of graphs from the $n\times n$ matrix with $0--1$ entries.   Note that the subgraph $\mathbb G_{(i,h)}$ here refers to all subgraphs that are homomorphic to $\mathbb G_{(i,h)}$,\footnote{In Graph theory, the notion of graph homomorphism is defined as follows: For a graph $G$, let $V(G)$ and $E(G)$ be the set of vertices and the set of edges of $G$, respectively. Let $G$ and $H$ be two graphs. A mapping $\varphi: V(G)\rightarrow V(H)$ is called {\it homomorphism} if $\varphi$ preserves edge adjacency, i.e., for  every edge $\{v,w\}\in E(G)$,  the edge $\{\varphi(v),\varphi(w)\}$ belongs to $E(H)$.} because players' identities do not matter in the definition of $\mathbb G_{(i,h)}$. Moreover, the first half of \Cref{SNdistr} also requires that $\mathbb G_{(i,h)}$ should be i.i.d. across players who are at least $2h$--step faraway from each other in the network. This condition generally holds in the random graph literature, since conditional on $\mathbb G_{(i,h)}$ and $\mathbb G_{(j,h)}$ do not overlap, the graph structure of $\mathbb G_{(i,h)}$ does not provide additional information on how $\mathbb G_{(j,h)}$ looks like.  Moreover, for the second half of \Cref{SNdistr}, the (conditional on the network connections) independence of $X_i$ is a strong assumption. In practice, positive statistic dependence across friends' demographic variables (e.g. age, education, race, etc.) has been emphasized in the sociology literature \citep[see e.g.][]{easley2010networks}, which is the so-called ``homophily'' phenomena.  For our asymptotic results to be established, this assumption could be relaxed to allow for some degree of dependence at the expense of longer proofs.\footnote{For instance, as is suggested by spatial autoregressive models, one could assume that $\mathbb X\equiv (X_1',\cdots,X_n')'$ takes a simultaneous autoregressive dependence structure: 
\[
\mathbb X = \Psi(\gamma_0)\cdot \mathbb X+\nu,
\]where $\Psi$ is an $n\times n$ weight matrix parametrized by a $q$-dimensional vector $\gamma_0$ such that diagonal elements of $\Psi$ are zeros and $\mathbb I_n-\Psi$ is non-singular. Moreover, $\nu\in \mathbb R^n$ is a vector of i.i.d. errors that are independent of $\mathbb S$ and $(\epsilon_1,\cdots,\epsilon_n)'$. Our asymptotic results established in \Cref{asymptotics,normality} still hold as long as for each $k\in\mathcal A$,
\[
\frac{1}{n} \sum_{i=1}^n \Big\{\sigma^*_{ik}(\mathbb S; \theta) \ln \sigma^*_{ik}(\mathbb S; c)-\mathbb E \left[\sigma^*_{ik}(\mathbb S; \theta) \ln \sigma^*_{ik}(\mathbb S; c)\right]\Big\}\overset{p}{\rightarrow} 0, \ \ \ \text{uniformly holds in } c\in\Theta.
\] Such a high level condition can be satisfied if the weight matrix is modeled as:  $\Psi_{j\ell}=\psi(\min\{\rho(j,\ell),\rho(\ell,j)\})$ where $\psi$ is a decreasing function that decays sufficiently fast (i.e., subjects to exponential decay). Moreover, it is also possible to allow the dependence  between $X_j$ and $X_\ell$, i.e., $\Psi_{j\ell}$, to depend not only on the network distance between $j$ and $\ell$, but also on the distance between $j$ (or $\ell$) and other players that connect (directly or indirectly) to both of them. See e.g. \cite{pinkse2002spatial}. 
}


\Cref{maximumneighbours} imposes restrictions on the number of best friends that a single individuals  could have. Note that  the upper bound $c_0$ does not depend on the network size $n$. This condition is crucial  for the $\sqrt n$--asymptotics of the proposed estimator when the data come from one single large network game: By \cref{maximumneighbours} and the NDD condition,  we can limit the dependence among all the observations. Similar assumptions can also be found in, e.g., \cite{morris2000contagion} for the contagion analysis in local--interaction games. 

It is worth pointing out that \Cref{maximumneighbours} is not imposed in most of the recent empirical network formation models. Such a restriction, however, can be easily accommodated in recent network formation models, e.g., \cite{christakis2010empirical} and \cite{mele2010structural}.  On the theoretic side of network formation, e.g., \cite{jackson1996strategic}  introduce a cost for players to maintain a direct friendship link, which similarly limits the number of direct links each individual could have. In our empirical application,  each student was allowed to nominate at most ten best friends. Such a restriction is reasonable in light of capacity constraints (e.g. time and/or effort) for students to make and keep their best--friends.  Therefore, a network formation model using the same dataset should impose such a restriction; otherwise the model cannot rationalize the data.

 
\Cref{choiceh}--(i) is intuitive for the approximation of  $\sigma^*_{i}(\mathbb S;\theta)$. Moreover, strengthening (i), \cref{choiceh}--(ii)  ensures the approximation error is negligible in the limiting distribution of the estimator. 

In \cref{JN}, $J_n(\theta)$ is the Fisher information matrix of the $n$--player game. \Cref{JN} requires that the Fisher information matrix has a non--degenerate limit when the network size goes to infinity. Note that the convergence of $J_n(\theta)$  is implied by  \Cref{stability} and  \cref{SNdistr}, since the distribution $\mathbb S_{(i,h)}$ convergence to a  limit for all $i$ as $n\rightarrow\infty$.  Hence,  in \cref{JN}, the essential restriction is the non--degeneracy of the limit. 

\begin{theorem}
\label{asymptotics}
Suppose that  \cref{extremevalue,para-id}, \ref{compact}-(i), \ref{SNdistr}, \ref{maximumneighbours}, and  \ref{choiceh}-(i) hold. Then $\hat{\theta}\overset{p}{\rightarrow}\theta$. 
\end{theorem}

Given the consistency of $\hat{\theta}$, we now establish its limiting distribution, which is shown to be identical to $\hat \theta_{MLE}$ under addition conditions.

\begin{theorem}
\label{normality}
Suppose that \cref{extremevalue,maximumneighbours,para-id,compact,stable2,SNdistr,choiceh,JN} hold. Then
$
\sqrt{n}(\hat{\theta}-\theta)\overset{d}{\rightarrow} N\left(0,J(\theta)^{-1}\right).
$ 
\end{theorem}
\noindent
Note that the infeasible likelihood function \eqref{likelihood} is indeed what ultimately gives the information equality in \Cref{normality}.  Furthermore, the limiting Fisher information matrix $J(\theta)$ can be consistently estimated by 
\[
\frac{1}{n}\sum_{i=1}^n \Big[\frac{\partial}{\partial \theta}  f^h_{i}(Y_i|\mathbb S; \hat{\theta}) \times \frac{\partial}{\partial \theta'}  f^h_{i}(Y_i|\mathbb S; \hat{\theta}) \Big],
\] where $f^h_{i}(Y_i|\mathbb S;  \theta)=\sum_{k\in A}\one (Y_i=k)\ln\sigma^h_{ik}(\mathbb S;\theta)$ and $
\frac{\partial}{\partial \theta} f^h_{i}(Y_i|\mathbb S; {\theta})=\sum_{k\in A}\frac{\one (Y_i=k)}{\sigma^h_{ik}(\mathbb S;\theta)}\times\frac{\partial}{\partial \theta} \sigma^h_{ik}(\mathbb S;\theta)$.

\subsubsection*{Remark}
It is a generic aspect of our asymptotic analysis that the size of the network goes to infinity, but the maximum number of friends each player should remain fixed (i.e. \cref {maximumneighbours}). Therefore, the collection of state variables $\{\mathbb S_{(i,h)}: i\leq n\}$  becomes an $m$--dependent sequence, where $m\leq c_0^{h+1}$, which is crucial for the $\sqrt n$--consistency of the estimator in the proof of \Cref{normality}. This aspect rules out the ``Small--World phenomenon'' \citep[see e.g.][]{watts1998collective}, often referred to as six degrees of separation \citep[see e.g.][]{guare1990six}. It should be noted that whether the network is a ``small--world'' is an empirical question that can be verified from the data.  In a small--world network, the asymptotic analysis should allow the (average) number of friends to increase with the size of the network.\footnote{\cite{watts1998collective} develop a small--world model by rewiring a regular network with $n\gg  Q_i\gg \ln n\gg 1$. }  It seems to be an intriguing challenge to consider  Small--World asymptotics. 


It is worth pointing out that it is possible to relax \cref {maximumneighbours}  to accommodate some ``intermediate'' case of the network structure at the expense of longer proofs.  For instance, consider a network where the maximum number of friends is not bounded from above, but the distribution of $Q_i$ is asymptotically stable (as the network size $n$ goes to infinity) with finite mean and variance.  Hence, there can be a few, but significant number of nodes with a lot of connections, which however does not render the network to a ``Small--World''.  In the following Monte Carlo experiments, we consider such a specification to examine the finite sample performance of our approximated maximum likelihood estimator.

\subsection{Monte Carlo Experiments}\label{MonteCarlo}
This section uses Monte Carlo to illustrate the finite sample performance of the proposed estimator. In particular, we
consider a binary game with payoff: $U_{i1}(Y_{-i},S_i,\epsilon_i)=X_i'\beta +\alpha\times \left[\frac{1}{Q_i}\sum_{j\in  F_ i}\one (Y_j=1)\right]+\epsilon_{i1}$, where $\alpha\in\mathbb R$ and   $X_i\in \mathbb R^2$.  

Moreover, we consider two representative networks: First, we consider the Circle network specified in \cite{salop1979monopolistic}, where $n$ players are equally spaced in a circle and each player has two friends. In the circle network, $Q_i=2$ for all players and the friendship relation between each pair of players is also symmetric. The second network  is a random network. For any $i\neq j$, we use a random variable $\vec{\ell}_{i,j}\in\{0,1,2,3\}$ to denote  ``no relationship'', ``$i$ is $j$'s friend, but not vice versa'', ``$j$ is $i$'s friend, but not vice versa'' and ``mutual friendship'', respectively. For $i\neq j$, $\vec{\ell}_{i,j}$ is drawn independently from the probability mass distribution $\big(1-\frac{4}{n},\frac{1}{n},\frac{1}{n},\frac{2}{n}\big)$. Moreover, set $\vec{\ell}_{ii}=0$ for all $i$. By definition, $Q_i=\sum_{j=1}^n\mathbb 1 (\vec{\ell}_{ij}\in\{2,4\})$, which conforms to a Binomial Distribution $B(n,3/n)$.  As $n$ goes to infinity, the mean of $Q_i$ remains constant and conforms to the Poisson(3) distribution asymptotically. 


Moreover, we take $X_{i1}\sim U(-0.5,0.5)$, $X_{i2}\sim N(0,1)$ and $X_{i1}\bot X_{i2}$. The results for the other distributional specifications of $X$ are qualitatively similar. Further, we set $\beta=(1,1)'$ which are invariant across all the experiments.  According to \cref{stable2}, we choose $\Theta_{\alpha}=[-1.99,1.99]$ and set the true parameter $\alpha=0$, $0.8$, and $1.6$, respectively. In particular, for $\alpha=0$, our setting is equivalent to the classical Logit model. 


We have performed experiments with the number of players $n=500$, $n=1000$ and $n=2000$. In each design, we first compute the unique BNE given the underlying parameter value, i.e., we solve the equilibrium by finding a fixed point  to (\ref{choiceprobability}).  With the (numerical) solution in hand, we are able to simulate the equilibrium decision made by each player.

Regarding estimation, it is crucial to choose the parameter $h\in\mathbb N$ according to the sample size $n$. Following \cref{choiceh}, we set $h=[\sqrt n/10]$, i.e. $h=2,3$ and $4$ with respect to the three choices of sample size. It is worth pointing out that  the computation time increases with $h$ in a non--linear pattern. For fixed $n$, we also investigate the performance of the proposed estimator under different choices of $h$. The results for different sample sizes are qualitatively similar and therefore we only report results for $n=1000$.  In addition, we perform $500$ replications to approximate the finite sample distribution of our estimator.

\Cref{table:2,table:3} report the finite sample performance of the proposed estimator under the different settings. The numbers in parentheses are the standard deviations. The estimator is consistent for all these designs and the standard deviation diminishes at the $\sqrt n$--rate as we increase the sample size. In \Cref{table:4}, we further investigate how the choice of $h$ affects the performance of $\hat\theta$. For $n=1000$, it shows that the approximation behaves well by using $h\geq 3$ and additional gains of accuracy are minor from choosing larger $h$.

\begin{table}[h]
\caption{ Finite sample performance:   $\beta=(1,1)$ and $(n,h)=(1000,3)$}
{\small
\begin{center}
\begin{tabular}{ccccccc}\hline
True value of $\alpha$       &Parameters  &  Circle Network  & Random Network\\ \hline
0                  & $\beta_1$   &1.0131      &1.0292\\
                   &                     &(0.2454)    &  (0.2493)\\
                    & $\beta_2$    &1.0036      &1.0058\\
                     &                    & (0.0826)    & (0.0833)\\
                     & $\alpha$      &0.0068      &0.0109\\         
                     &                     &(0.1326)  &  (0.1402)  \\\hline                           
0.8              & $\beta_1$    &1.0018      &          1.0204\\
                   &                     &(0.2468)    &          (0.2557)\\
                   & $\beta_2$    & 1.0091     &         1.0060\\
                   &                     & (0.0833)   &        (0.0834)\\
                   & $\alpha$      &  0.8066    &         0.8023\\
                   &                    &(0.1042)  &         (0.1114) \\\hline           
1.6              & $\beta_1$    &1.0059     &        1.0179\\
                   &                     & (0.2464) &         (0.2721)\\
                   &$\beta_2$     &1.0008  &        1.0064\\
                   &                    &(0.0849)  &       (0.0839)\\
                  & $\alpha$      & 1.6256   &      1.6169    \\
                  &                    & (0.0950) &      (0.0930)\\ \hline
\end{tabular}
\end{center}}
\label{table:2}
\end{table}

\begin{table}[h]
\caption{ Finite sample performance of $\hat\alpha$}
{\small
\begin{center}
\begin{tabular}{ccccccc}\hline
True value of $\alpha$ & Sample size      &  Circle Network & Random Network\\ \hline
0                 &  500    &  0.0030                   & 0.0033\\
                   &           &         (0.1954)               &(0.2004) \\
                   & 1,000  &0.0068       &0.0109\\
                   &            &(0.1326)     &(0.1402)\\
                   & 2,000  &0.0044       &0.0022\\
                   &            &(0.0962)      &(0.0968)\\\hline
0.8              &500      &    0.8032                   &0.8048\\
                   &            &     (0.1570)                  &(0.1469)\\
                   & 1,000  & 0.8066          & 0.8023\\
                   &            &  (0.1042)       & (0.1114)\\
                   & 2,000  & 0.8036     &0.7964\\
                   &            & (0.0714)   &(0.0716)\\\hline
1.6              &500       &  1.6254                     &1.6776\\
                   &            &   (0.1282)          &(0.1398)\\
                   & 1,000  &  1.6256           &1.6169\\
                   &            &  (0.0950)        &(0.0930)\\
                   & 2,000  & 1.6072 & 1.6064\\
                   &            &(0.0660) & (0.0659)\\\hline
\end{tabular}

\end{center}}
*Note that $h=2, 3, 4$ for $n=500, 1000$ and $2000$, respectively.
\label{table:3}
\end{table}

\begin{table}[h]
\caption{ Finite sample performance of $\hat\theta$ at different $h$ ($n=1000$, $\alpha=0.8$)}
{\small
\begin{center}
\begin{tabular}{lccccccc}\hline
                 & Parameters & $h=0$ & $1$           & 2             &3           &4               \\\hline
Circle Network       & $\beta_1$    &0.9790   &1.0121    &1.0155   &1.0157  &1.0157    \\
                              &                    &(0.2501) &(0.2448) &(0.2461) & (0.2462)& (0.2462) \\                          
                              &$\beta_2$    &0.9627     &0.9967   & 1.0002   &1.0004 &1.0004   \\
                              &                    &(0.0821)  &(0.0845) & (0.0848) &(0.0849)&(0.0849)  \\
                              &$\alpha$       &n/a    &0.8560   &0.8014    &0.7974  &0.7972    \\
                              &                     &n/a  &(0.1118) & (0.0996)&(0.0986) &(0.0984)  \\ \hline      
Random Network   &  $\beta_1$   &0.9649& 1.0063   &    1.0094     & 1.0098& 1.0098 \\
                              &                     &(0.2568)& (0.2575)  & (0.2584)   &(0.2585)&(0.2585)\\
                              &$\beta_2$     &0.9614&    0.9990  &  1.0023      & 1.0026 &1.0026  \\
                              &                     &(0.0823)&    (0.0824)  & (0.0825)   &(0.0825)&(0.0825)\\                   
                              &$\alpha$       &n/a& 0.8957     & 0.8068         &0.7979&0.7968\\
                             &                     &n/a&   (0.1289)    &    (0.1063) &(0.1033)&(0.1028) \\\hline           
\end{tabular}
\end{center}
 }

\label{table:4}
\end{table}

\subsection {Empirical results for peer effects on college attendance}
We now apply our method to estimate peer effects on high school students' college attendance decisions. The specification of the payoff function is the same as the one used in our Monte Carlo experiments.

\Cref{table6} presents our estimation results. We also provide results using the pseudo MLE for comparison. The difference reflects the bias due to the misspecification of social interactions. Note that AMLE(h) refers to the approximated MLE with the parameter value $h$ and the pseudo MLE is equivalent to AMLE(0).  From \Cref{table6}, the approximation of the equilibrium is sufficiently good for $h\geq 2$. So we can use AMLE(2) as our estimates. It is worth pointing out that the estimates of peer effects satisfy \cref{smallstrategyeffect}.

The second column of \Cref{table6} contains the corresponding estimates of the pseudo MLE, which has been typically adopted in the empirical analysis on college attendance. Given  the pseudo MLE estimates, the most striking difference of our estimates (i.e. AMLE(2) in the fourth column) is  that the peer effects coefficient  is significant at the 5\% level, while the pseudo MLE implicitly sets it to be zero. Therefore, the ignorance of peer effects in the empirical analysis on college attendance results in biased estimates, which can be corrected by  increasing $h$ from $0$ to $2$.

In \Cref{table6}, most of coefficients estimates are significant at the 10\% significance level. Regarding race, the coefficients of American Indian, Asian  and Black are insignificant, this is simply due to the fact that all these three categories have only a few observations in the sample. Moreover, due to missing data issue on parents' education, one would expect noisy estimates for the parents' education coefficients. 

Our pseudo MLE estimates are qualitatively similar to those empirical results in \cite{light2002bakke} who estimate racial effects on college attendance with a Probit model by using the data from the 1979 National Longitudinal Survey of Youth (NLSY79), which consists a sample of respondents born in 1957--1964. In particular, whites are less likely than minorities to attend college, given other determinants of college attendance are held constant. For such a comparison, note that peer effects are not considered in \cite{light2002bakke}. Our pseudo MLE results are also consistent with other early empirical evidence on college attendance. See e.g. \cite{fuller1982new}.\footnote{\cite{fuller1982new} use the 1972 National Longitudinal Study of the High School Class (NLSS72).}

Peer effects estimates provided by AMLE(2)  are related to those empirical results in  \cite{calvo2009peer}, who also use the Add Health data to study peer effects on school performance index. In particular, they specify a linear equation system for network--based social interactions and obtain statistically significant peer effects estimates of similar magnitude (i.e., 0.5505 with a standard error 0.1247). Moreover, \cite{gaviria2001school} and \cite{kawaguchi2004peer} use the National Education Longitudinal Study (NELS) dataset and the National Longitudinal Survey Youth 97 (NLSY97) dataset, respectively, to study  peer effects on youth behaviors of high school students, e.g., drug use, alcohol drinking, cigarette smoking, church attendance and dropping out. Their empirical results also provide evidence for significant peer effects of similar magnitude to our estimates. For example, consider a typical student in our sample whose covariates take the mean values in \Cref{table5}. Suppose all her friends shift their college attendance probabilities together from 0\% to 10\%,  then her college attendance probability would increase about 1.52\% (namely, from 37.93\% to 39.45\%). Similarly, if all her friends' college attendance probabilities shift jointly from 0\% to 50\%, then it would yield an increase of 11.83\%.\footnote{In a study of tenth graders' substance use,  \cite{gaviria2001school}'s estimates imply, for example, that moving a typical teenager from a school where none of his classmates use drugs to one where half use drugs would increase the probability  by approximately 13\%. Similar experiments would yield increases in the corresponding probabilities of 9\% for alcohol use, 8\% for cigarette smoking, 11\% for church attendance, and 8\% for dropping out of school. Moreover, \cite{kawaguchi2004peer} show that if a teenager's perception of the percentage of his/her peers who use a substance (i.e. marijuana, alcohol, or tobacco) increases by 10\%, the probability that he/she will
use the substance increases from 1.4\% to  2.6\%.}

\begin{table}[h]
\small
\caption{Estimation Results}
\label{table6}
\begin{center}
\begin{tabular}{lccccc}
\hline\hline
Variable                     & Pseudo MLE         & AMLE(1)   &AMLE(2)&AMLE(3)   &AMLE(4)\\\hline
Age                            &-0.140*       &   -0.135*     &-0.135*    &-0.135*      &-0.135* \\
                                  & (0.076)       &(0.076)       & (0.076)    & (0.076)   & (0.076)\\
Female                      &-0.028       &-0.038      &-0.035    &-0.034      &-0.034 \\
                                  & (0.171)     & (0.171)      &(0.171)   &(0.171)     &(0.171)\\
Household Income    &0.150**     &  0.134**     &0.134**     &0.134**       &0.134**   \\
                                  &(0.042)      & (0.043)     &(0.043) &(0.043)      &(0.043)\\ 
Mother's Education      &0.066        &  0.064        &0.064   &0.064       &0.064\\
                                  &(0.052)     &  (0.053)      &(0.053)  &(0.053)      &(0.053)\\
Father's Education      &0.033         & 0.035         &0.036     &0.036       &0.036 \\
                                 &(0.029)       &  (0.029)     &(0.029)  &(0.029)      &(0.029)\\
Overall GPA             &1.749**     &   1.714**        &1.717**     &1.717**       &1.717** \\
                                 &(0.147)      &   (0.148)      &(0.148)   &(0.148)       &(0.148)\\
American Indian      &-0.559       &  -0.575       &-0.574     &-0.574       &-0.574 \\
                                 &(0.418)     &   (0.423)     &(0.423)    &(0.423)      &(0.423)\\
Asian                        &-0.050        &   0.035       & 0.043     &0.043         &0.043\\
                                 &(0.428)     &  (0.435)      & (0.435)   &(0.435)       &(0.435)\\
Black                        &0.206        & 0.351          &0.363     & 0.364      & 0.364\\
                                &(0.455)      & (0.466)       & (0.467)   &(0.467)       &(0.467)\\
Hispanic                  &0.891**       & 1.043**          & 1.051**      &1.052**       &1.052**\\
                               &(0.223)       &(0.233)        & (0.234)  &(0.234)       &(0.234)\\
White                       &-0.703*       &-0.718*         &-0.717*   &-0.718*       &-0.718* \\
                                &(0.393)      &(0.401)         & (0.401)   & (0.401)      & (0.401)\\
Other Race              & -1.024**    &-1.096**       &   -1.097**  &  -1.098**     &  -1.098**\\
                                & (0.422)    &(0.430)         & (0.430)   &(0.430)     &(0.430)\\
Constant                 &    -2.680*    &      -2.795*    &  -2.806*    &  -2.806*      &  -2.806*\\
                               &  (1.441)       &      (1.445)  & (1.446)     & (1.446)       & (1.446)\\
Peer Effects             & \ \ ---          &0.657**          &0.642**      &0.640**        &0.640**\\
                                &\ \ ---           &(0.297)          & (0.286) &(0.285)      &(0.285)\\ \hline
LogLikelihood          & -437.537    & -435.063    &-434.988&-434.990&-434.990\\                           
\hline
\multicolumn{6}{l}{\footnotesize{*\ significant at 10\% level.}}\\
\multicolumn{6}{l}{\footnotesize{**\ significant at 5\% level.}}\\
\end{tabular}
\end{center}
\end{table}

\section{Conclusion}
This paper provides a structural approach to study social interactions in a large network. Our benchmark model assumes that individuals are affected by their friends only but all individuals are connected to each other directly or indirectly in a single network. By restricting the strength of interactions among friends, we establish the existence, uniqueness of the equilibrium and a NDD condition. We further establish the semiparametric identification of the model and propose a computationally feasible and novel estimation procedure. The classic MLE method  developed in  single--agent binary response models is naturally nested in our approach.

An important extension of the benchmark model is to allow for interdependence between a pair of friends' private information.  Individuals tend to bond with similar others as their friends. In sociology, such a phenomena is called ``homophily''; see e.g. \cite{easley2010networks}. Homophily leads to friendship between people with similar characteristics (age, education, race, etc.) and with positively correlated types. The former can be directly observed from the data. To identify the latter is more challenging to the researcher.  In a discrete game with a (small) fixed number of players,  \cite{liu2012rationalization} establish the nonparametric identification of homophily in a context of discrete game. Identification and estimation of homophily in a large network game is an important extension. 

Allowing for possible endogeneity of the network is another important research question in the study of large--network social interactions. Being popular in a hight school network might be associated with a possible high draw of payoff shocks for college attendance. Part of the problem could be addressed by taking into account the network formation in the first stage, see e.g. \cite{christakis2010empirical}, \cite{mele2010structural}, \cite{badev2013discrete} \cite{leung2014random} and \cite{menzel2015strategic}. In this regards, our identification and estimation results are useful for the second stage analysis of social interactions in the subgame. In a large network game, however, difficulties arise when each player has a small opportunity set, relative to the large network size, of players to meet with, and more importantly, such opportunity sets are not observed in the dataset. For the majority of pairs of distinct individuals, it is unclear whether an unconnected link is due to the lack of opportunity, or players' unfavorable desire for such a connection. 

As a matter of fact, our results go well beyond the local interaction studied here as they can be generalized to more general social interaction games. For instance, one can consider that each player interacts directly with her friends, friends of friends, etc. In particular, the payoff function can be generalized as follows: for choosing an action $k\in A$, 
 \begin{equation*}
 U_{ik}( Y_{-i},S_i, \epsilon_i)=\beta_k(X_i)+\sum_{j\neq i}\alpha_k( Y_j,d_{ij}, X_i, Q_i)+\epsilon_{ik},
 \end{equation*}where $d_{ij}$ is the network distance from $j$ to $i$. By such an extension, the interaction term $\alpha_k( Y_j,d_{ij}, X_i, Q_i)$ depends on player $j$'s choice as well as their network distance. In (\ref{payoff1}), direct interactions $\alpha_k$  have been set to zero for all $j\not\in F_i$.  By a similar argument, our uniqueness and NDD condition of the equilibrium can be established. A major difficulty in developing nonparametric identification and estimation, however, is to consider a model with an increasing parameter space, since the support of $d_{ij}$ expands with the size of the network. Though significant progress has been made in the regression context \citep[see, e.g.,][]{belloni2011high}, the different nature of the structural analysis calls for further work.

\clearpage

\bibliographystyle{econometrica}	
\bibliography{bibles}		

\clearpage

\appendix
\section{Equilibrium uniqueness and network stability}
\subsection{Proof of Lemma \ref{uniqueness} } \label{uniquenessproof}

Fix $n$ and $\mathbb S=s$. We prove by contradiction. Suppose there are two BNEs, denoted by $\{\sigma^*_i: i=1,\cdots,n\}$ and $\{\sigma^\dag_i: i=1,\cdots,n\}$ respectively. For notational simplicity, throughout we suppress their dependence on $\mathbb S$ and $\theta$.

For any choice probability profile $(\sigma_1,\cdots,\sigma_n)$, where $\sigma_i$ is a (K+1)--choice probability distributions, let 
\begin{equation}
\label{gamma}
\Gamma_{ik}\left(s_i, \{\sigma_{j}: j\in F_i\}\right)=\frac{\exp\left[\beta_k(x_i)+\sum_{\ell=0}^K\left\{\alpha_k(\ell,x_i,q_i)\sum_{j\in F_i}\sigma_{j\ell}\right\}\right]}{1+\sum_{\ell'=1}^K\exp\left[\beta_{\ell'}(x_i)+\sum_{\ell=0}^K\left\{\alpha_{\ell'}(\ell,x_i,q_i)\sum_{j\in F_i}\sigma_{j\ell}\right\}\right]}.
\end{equation} Let further $\Gamma_i\left(s_i, \{\sigma_{j}: j\in F_i\}\right)=\big(\Gamma_{i0}(s_i, \{\sigma_{j}: j\in F_i\}),\cdots, \Gamma_{iK}(s_i, \{\sigma_{j}: j\in F_i\})\big)'$. By \cref{choiceprobability}, we have $\sigma^*_{i}=\Gamma_{i}(s_i,\{\sigma^*_j: j\in F_i\})$ and $\sigma^\dag_{i}=\Gamma_{i}(s_i,\{\sigma^\dag_j: j\in F_i\})$ for all $i\in N$.

Therefore, for any $i\in N$, 
\[
\sigma^*_i-\sigma^\dag_i
=\Gamma_{i}\big(s_i,\{\sigma^*_j: j\in F_i\}\big)-\Gamma_{i}\big(s_i,\{\sigma^\dag_j:j\in F_i\}\big)
=\sum_{j\in F_i}\sum_{\ell\in A}\frac{\partial \Gamma_i(s_i,\{\tilde \sigma_j:j\in F_i\})}{\partial \sigma_{j\ell}}\cdot (\sigma^*_{j\ell}-\sigma^\dag_{j\ell})
\]
where $\{\tilde\sigma_j: j\in F_i\}$ is a vector between $\{\sigma^*_j: j\in F_i\}$ and $\{\sigma^\dag_j: j\in F_i\}$. By the definition of $\Gamma_{ik}$, we have 
\begin{multline*}
 \frac{\partial \ln \Gamma_{ik}}{\partial \sigma_{j\ell}}
=\alpha_{k}(\ell,x_i,q_i)-\sum_{\ell'=1}^K\Gamma_{i\ell'}\cdot \alpha_{\ell'}(\ell,x_i,q_i)\\
=\sum_{\ell'=0}^K\Gamma_{i\ell'}\cdot \alpha_{k}(\ell,x_i,q_i)-\sum_{\ell'=0}^K\Gamma_{i\ell'}\cdot \alpha_{\ell'}(\ell,x_i,q_i)
\end{multline*}where the last step is because: (i) $\sum_{\ell'=0}^K\Gamma_{i\ell'}=1$; (ii) $\alpha_{0}(\ell,x,q)=0$. It follows that 
\[
\frac{\partial\Gamma_{ik}}{\partial\sigma_{j\ell}}=\Gamma_{ik}\sum_{k'\neq k}\left[\Gamma_{ik'}\cdot\left\{\alpha_k(\ell,x_i,q_i)-\alpha_{k'}(\ell,x_i,q_i)\right\}\right].
\] Therefore,
\[
 \sum_{k\in A}\left|\frac{\partial \Gamma_{ik}}{\partial\sigma_{j\ell}}\right|\leq {\Delta}^*(x_i,q_i)\cdot\sum_{k\in A}\left[\Gamma_{ik}\left(1-\Gamma_{ik}\right)\right]\leq {\Delta}^*(x_i,q_i)\cdot \frac{ K}{K+1}.
 \] where ${\Delta}^*(x,q)\equiv\max_{k,\ell,m\in \mathcal{A}}\left|\alpha_k(\ell, x,q)-\alpha_m(\ell,x,q)\right|$ and the last step comes from the fact that (i) $0\leq\Gamma_{ik}\leq 1$; (ii) $\sum_{k=0}^K \Gamma_{ik}=1$. Hence,
\begin{multline*}
\|\sigma^*_i-\sigma^\dag_i\|_1=\sum_{k\in A}\Big|\sum_{j\in F_i}\sum_{\ell\in A}\frac{\partial \Gamma_{ik}(s_i,\{\tilde \sigma_j:j\in F_i\})}{\partial \sigma_{j\ell}}\cdot (\sigma^*_{j\ell}-\sigma^\dag_{j\ell})\Big|\\
\leq  \sum_{j\in F_i}\sum_{\ell\in A} \Bigg\{\left|\sigma^*_{j\ell}-\sigma^\dag_{j\ell}\right|\cdot \sum_{k\in A} \Big|\frac{\partial \Gamma_{ik}(s_i,\{\tilde \sigma_j:j\in F_i\})}{\partial \sigma_{j\ell}}\Big|\Bigg\}\\
\leq \Delta^*(x_i,q_i)\cdot \frac{K}{K+1} \cdot \sum_{j\in F_i}\sum_{\ell\in A} \left|\sigma^*_{j\ell}-\sigma^\dag_{j\ell}\right|\\
\leq \Delta^*(x_i,q_i)\cdot \frac{K}{K+1} \cdot q_i\cdot \max_{j\in F_i} \|\sigma^*_{j}-\sigma^\dag_{j}\|_1\leq \lambda\cdot  \max_{j\in F_i} \|\sigma^*_{j}-\sigma^\dag_{j}\|_1.
\end{multline*}

Therefore, 
\[
\max_{i\in N} \|\sigma^*_i-\sigma^\dag_i\|_1\leq \lambda\cdot  \max_{i\in N} \max_{j\in F_i} \|\sigma^*_{j}-\sigma^\dag_{j}\|_1\leq \lambda \cdot \max_{j\in N} \|\sigma^*_{j}-\sigma^\dag_{j}\|_1
\]which leads to contradiction by $\lambda<1$ under \cref{smallstrategyeffect}.
\qed

\subsection{Proof of Lemma \ref{stability} } \label{proof_stability}
This lemma is shown by mathematical induction. Fix arbitrarily $n,h\in\mathbb N$ and $s,s'\in\mathbb S$ such that $s_{(i,h)}=s'_{(i,h)}$.  

First, for all $j\in N_{(i,h)}$,  we have $s_j=s'_j$. We now derive $\sigma^*_j(s ; \theta)-\sigma^*_j(s';\theta)$ using Taylor expansion, i.e.,
\[
\sigma^*_j(s'; \theta)-\sigma^*_j(s; \theta)=\sum_{j\in F_i}\sum_{\ell\in A}\frac{\partial \Gamma_j(s_j,\{\tilde \sigma_{j'}:j'\in F_j\})}{\partial \sigma_{j'\ell}}\cdot (\sigma^*_{j'\ell}(s'; \theta)-\sigma^\dag_{j'\ell}(s; \theta))
\]where $\{\tilde\sigma_{j'}: j'\in F_j\}$ is a vector between $\{\sigma^*_{j'}(s; \theta): j\in F_i\}$ and $\{\sigma^*_{j'}(s'; \theta): j\in F_i\}$. By a similar argument to the proof of Lemma \ref{uniqueness}, we have
\begin{multline*}
 \|\sigma^*_j(s ; \theta)-\sigma^*_j(s'; \theta)\|_1\leq \lambda\cdot \max_{j'\in F_j}  \|\sigma^*_{j'}(s; \theta)-\sigma^*_{j'}(s'; \theta)\|_1\\
 \leq \lambda\cdot \max_{j'\in F_j} \{ \|\sigma^*_{j'}(s; \theta)\|_1+\|\sigma^*_{j'}(s'; \theta)\|_1\}= 2\lambda,
\end{multline*}where the last inequality comes from the triangular inequality.  Because for all $j\in N_{(i,h-1)}$,  any friend $j'$ of $j$ belongs to $N_{(i,h)}$, then 
\[
 \|\sigma^*_j(s ; \theta)-\sigma^*_j(s'; \theta)\|_1\leq \lambda^2\cdot \max_{j''\in F_{j'}, j'\in F_i}  \|\sigma^*_{j''}(s ; \theta)-\sigma^*_{j''}(s'; \theta)\|_1\leq 2\lambda^2.
\]By induction, for all $j\in N_{(i,h-q)}$ where $q\leq h$,  there is 
\[
 \|\sigma^*_j(s ; \theta)-\sigma^*_j(s'; \theta)\|_1\leq 2\lambda^{q+1}.
\]
Hence, for any $q\leq h$, we have
\[
\max_{j\in N_{(i,h-q)}} \|\sigma^*_j(s ; \theta)-\sigma^*_j(s'; \theta)\|_1\leq 2\lambda^{q+1}.
\]
Because $i\in N_{(i,0)}$, then $ \|\sigma^*_i(s; \theta)-\sigma^*_i(s'; \theta)\|_1\leq 2\lambda^{h+1}$. By  \cref{smallstrategyeffect}, $2\lambda^{h+1}\downarrow 0$ as $h\rightarrow \infty$.
\qed

\subsection{Proof of Lemma \ref{pointidentification} } 
\label{proof_pointidentification}
First, by \cref{payoff_normalization}, (\ref{logchoiceprobability}) can be rewritten as
\[
\delta_{ik}(\mathbb S)=\varphi'_i(\mathbb S)\times  \Bigg(\begin{array}{c}\beta_k(X_i) \\ \alpha_k(X_i,Q_i)  \end{array}\Bigg).
\]
We further multiply by $\varphi_i(\mathbb S)$ on both sides and obtain
\[
\varphi_i(\mathbb S)\times \delta_{ik}(\mathbb S)=\varphi_i(\mathbb S)\times \varphi'_i(\mathbb S)\times  \Bigg(\begin{array}{c}\beta_k(X_i) \\ \alpha_k(X_i,Q_i)  \end{array}\Bigg).
 \]
Moreover, we take conditional expectation on both sides given $X_i=x$ and $Q_i=q$:
\[
\mathbb E[\varphi_i(\mathbb S)\times \delta_{ik}(\mathbb S)|X_i=x,Q_i=q]=\mathbb E[\varphi_i(\mathbb S)\times \varphi'_i(\mathbb S)|X_i=x,Q_i=q]\times  \Bigg(\begin{array}{c}\beta_k(x) \\ \alpha_k(x,q)  \end{array}\Bigg)
 \]from which we invert the coefficients vector $(\beta_k(x), \alpha'_k(x,q))'$.\qed

\section{Asymptotic properties under parametric setting}
For any $c\in\Theta$, let $L_n(c)=\frac{1}{n} \sum_{i=1}^n \sum_{k\in A} \mathbb E \left[\sigma^*_{ik}(\mathbb S; \theta) \ln \sigma^*_{ik}(\mathbb S; c)\right]$. For arbitrary $\epsilon>0$, let ${B}_\epsilon(\theta)$ be an open ball centered at $\theta$ with $\epsilon$ radius in the space $\Theta$.

\subsection{Proof of Theorem \ref{asymptotics}}\label{proofasymptotics} 
 
By Lemma \ref{consistency}, it suffices to check the conditions (i) -- (iii) in the lemma. By the identification argument and \cref{para-id}, condition (i) holds.  Moreover, condition (iii) also holds by Lemma \ref{lemma3}. Hence, it suffices to verify condition (ii), i.e. $\sup_{c\in\Theta}\big|\hat{L}(c)-L_n(c)\big|\overset{p}{\rightarrow} 0$.

By Lemmas \ref{bounded} and \ref{c5}, $ \sum_{k=0}^K \one (Y_i=k) \ln \sigma^*_{ik}(\mathbb S;\cdot)$ is bounded and continuous on $\Theta$. Since $\Theta$ is compact,  then $\mathscr{F}_n= \left\{\sum_{k\in A} \one (Y_i=k) \ln \sigma^*_{ik}(\mathbb S; c): c\in\Theta\right\}$ can be covered by a finite number of $\epsilon$--brackets. To apply the classical Glivenko-Cantelli argument, it suffices to show the point--wise law of large number, i.e. for any $c\in\Theta$, $\hat{L}(c)-L_n(c)\overset{p}{\rightarrow} 0$.

We pick an integer $d_n \propto 0.5 \ln n/\ln c_0$. Clearly, $d_n\rightarrow\infty$ as $n\rightarrow\infty$. Then  we have
\begin{multline}
\label{decom}
\hat{L}(c)-L_n(c)=\frac{1}{n} \sum_{i=1}^n \sum_{k\in A} \big\{\one (Y_i=k)-\sigma^*_{ik}(\mathbb S; \theta)\big\} \ln \sigma^*_{ik}(\mathbb S; c)\\
+\frac{1}{n} \sum_{i=1}^n \sum_{k\in A} \big\{\sigma^*_{ik}(\mathbb S; \theta) \ln \sigma^*_{ik}(\mathbb S; c)-\sigma^{d_n}_{ik}(\mathbb S; \theta) \ln \sigma^{d_n}_{ik}(\mathbb S; c)\big\}\ \ \ \ \ \ \  \\
+\frac{1}{n} \sum_{i=1}^n \sum_{k\in A} \left\{\sigma^{d_n}_{ik}(\mathbb S; \theta) \ln \sigma^{d_n}_{ik}(\mathbb S; c)-\mathbb E \left[\sigma^{d_n}_{ik}(\mathbb S; \theta) \ln \sigma^{d_n}_{ik}(\mathbb S; c)\right]\right\}\\
+\frac{1}{n} \sum_{i=1}^n \sum_{k\in A} \left\{\mathbb E\left[ \sigma^{d_n}_{ik}(\mathbb S; \theta) \ln \sigma^{d_n}_{ik}(\mathbb S; c)\right]-\mathbb E \left[\sigma^*_{ik}(\mathbb S; \theta) \ln \sigma^*_{ik}(\mathbb S; c)\right]\right\}.
\end{multline}
For the first term of right--hand side in \cref{decom}, we have
\begin{multline*}
\mathbb{E}\left\{\Big[\frac{1}{n} \sum_{i=1}^n \sum_{k\in A} \left(\one (Y_i=k)-\sigma^*_{ik}(\mathbb S; \theta)\right) \ln \sigma^*_{ik}(\mathbb S; c)\Big]^2\Big|\mathbb S\right\}\\
=\frac{1}{n^2}\sum_{i=1}^n\mathbb{E}\left\{\Big[ \sum_{k\in A} \left(\one (Y_i=k)-\sigma^*_{ik}(\mathbb S; \theta)\right) \ln \sigma^*_{ik}(\mathbb S; c)\Big]^2\Big|\mathbb S\right\}
\leq \frac{1}{n} (K+1)^2(\ln \sigma_0)^2\rightarrow 0
\end{multline*}where the first step is because of the reasons that $Y_i$ is conditionally independent given $\mathbb S$ and that $\mathbb E (Y_i|\mathbb S)=\sigma^*_{ik}(\mathbb S; \theta)$, and the last inequality is due to the fact: $\ln \sigma_0\leq \big(\one (Y_i=k)-\sigma^*_{ik}(\mathbb S; \theta)\big) \ln \sigma^*_{ik}(\mathbb S; c)\leq -\ln \sigma_0$ under Lemma \ref{bounded}. 

Next, for the second term of RHS in \cref{decom}, note that
\begin{multline*}
\mathbb E\left|\sigma^*_{ik}(\mathbb S; \theta) \ln \sigma^*_{ik}(\mathbb S; c)-\sigma^{d_n}_{ik}(\mathbb S; \theta) \ln \sigma^{d_n}_{ik}(\mathbb S; c)\right|\\
\leq \mathbb E\left[\big|\sigma^*_{ik}(\mathbb S; \theta) -\sigma^{d_n}_{ik}(\mathbb S;\theta)\big| \cdot \left|\ln \sigma^*_{ik}(\mathbb S; c)\right|\right]+\mathbb E\left[\big|\sigma^{d_n}_{ik}(\mathbb S;\theta)\big|\cdot \big|\ln \sigma^*_{ik}(\mathbb S; c)- \ln \sigma^{d_n}_{ik}(\mathbb S; c)\big|\right]\\
\leq -\ln\sigma_0\cdot \mathbb E \big|\sigma^*_{ik}(\mathbb S; \theta)-\sigma^{d_n}_{ik}(\mathbb S;\theta)\big|+\frac{1}{\sigma_0}\cdot \mathbb E \big|\sigma^*_{ik}(\mathbb S; c)-\sigma^{d_n}_{ik}(\mathbb S; c)\big|\rightarrow 0.
\end{multline*} Similarly, we can show that the last term  in \cref{decom} is also $o_p(1)$.

Therefore, it suffices to show that the third term of RHS in \cref{decom} is also $o_p(1)$.  Note that
\begin{multline*}
\mathbb E\Big\{\frac{1}{n} \sum_{i=1}^n \sum_{k\in A} \big[\sigma^{d_n}_{ik}(\mathbb S; \theta) \ln \sigma^{d_n}_{ik}(\mathbb S; c)-\mathbb E \sigma^{d_n}_{ik}(\mathbb S; \theta) \ln \sigma^{d_n}_{ik}(\mathbb S; c)\big]\Big\}^2\\
=\frac{1}{n^2} \sum_{i,j=1}^n \text{Cov}\left(\sum_{k\in A} \sigma^{d_n}_{ik}(\mathbb S; \theta) \ln \sigma^{d_n}_{ik}(\mathbb S; c), \sum_{k\in A} \sigma^{d_n}_{jk}(\mathbb S; \theta) \ln \sigma^{d_n}_{jk}(\mathbb S; c)\right).
\end{multline*}
By definition and \cref{SNdistr}, $\sigma^{d_n}_{i}(\mathbb S; \theta) $ is independent of $  \sigma^{d_n}_{j}(\mathbb S; \theta)$ if there does not exist a player $m\in N_{(i,d_n)}\bigcap N_{(j,d_n)}$. By \cref{maximumneighbours}, there are at most $n\cdot (1+c_0+\cdots c^{d_n}_0)\leq n c^{d_n+1}_0$ pair of $(i,j)$ such that $\sigma^{d_n}_{i}(\mathbb S; \theta) $ and $ \sigma^{d_n}_{j}(\mathbb S; \theta)$ are dependent of each other. Moreover, for any $i$ and $j$, 
\begin{multline*}
2\text{Cov}\Big(\sum_{k\in A} \sigma^{d_n}_{ik}(\mathbb S; \theta) \ln \sigma^{d_n}_{ik}(\mathbb S; c), \sum_{k\in A} \sigma^{d_n}_{jk}(\mathbb S; \theta) \ln \sigma^{d_n}_{jk}(\mathbb S; c)\Big)\\
\leq \mathbb E \Big(\sum_{k\in A} \sigma^{d_n}_{ik}(\mathbb S; \theta) \ln \sigma^{d_n}_{ik}(\mathbb S; c)\Big)^2+\mathbb E\Big(\sum_{k\in A} \sigma^{d_n}_{jk}(\mathbb S; \theta) \ln \sigma^{d_n}_{jk}(\mathbb S; c)\Big)^2 
\leq 2 (1+K)^2(\ln\sigma_0)^2.
\end{multline*}
Therefore, 
\begin{multline*}
\mathbb E\Big\{\frac{1}{n} \sum_{i=1}^n \sum_{k\in A} \big\{\sigma^{d_n}_{ik}(\mathbb S; \theta) \ln \sigma^{d_n}_{ik}(\mathbb S; c)-\mathbb E \sigma^{d_n}_{ik}(\mathbb S; \theta) \ln \sigma^{d_n}_{ik}(\mathbb S; c)\big\}\Big\}^2\\
\leq \frac{1}{n^2}\cdot  nc_0^{d_n+1} 2 (1+K)^2(\ln\sigma_0)^2\propto \frac{1}{\sqrt n}2c_0(1+K)^2(\ln\sigma_0)^2\rightarrow 0. \qed
\end{multline*}

\begin{lemma}
 \label{consistency}
 Suppose (i) $ \limsup_{n\rightarrow\infty}\sup_{c\not\in  B_\epsilon(\theta)}\left({L_n}(c)-L_n(\theta)\right)<0$ holds for any $\epsilon>0$; (ii)  $\hat{L}_n$ converges uniformly in probability to $L_n$, i.e. $ \sup_{c\in\Theta}\left|\hat{L}_n(c)-L_n(c)\right|\overset{p}{\rightarrow} 0$;
  (iii) $\hat{L}_n(\hat{\theta})\geq \sup_{c\in\Theta}\hat{L}_n(c)-o_p(1)$. Then $\widehat{\theta}\overset{p}{\rightarrow} \theta$.
 \begin{proof} 
To prove the lemma, we modify the proofs in \cite{newey1994large}, Theorem 2.1. Note that the objective function $L_n(\cdot)$ in our case  depends on $n$, and it converges to a limit as $n$ goes to infinity.  By (ii) and (iii), with probability approaching one (w.p.a.1),
\[
L_n(\hat{\theta})>\hat{L}_n(\hat{\theta})-\eta/3> \hat{L}_n({\theta})-2\eta/3 > L_n({\theta})-\eta, \ \ \forall \eta>0.
\]

Then, for any $\epsilon>0$, choose $\eta=-\frac{1}{2}\limsup_{n\rightarrow\infty}\sup_{c\not\in  B_\epsilon(\theta)}\left({L_n}(c)-L_n(\theta)\right)>0$. It follows that w.p.a.1,  
\[
L_n(\hat{\theta})-L_n({\theta})>\frac{1}{2}\limsup_{n\rightarrow \infty}\sup_{c\not\in  B_\epsilon(\theta)}\left({L_n}(c)-L_n(\theta)\right).
\]
Because for sufficient large $n$, 
\begin{multline*}
\sup_{c\not\in  B_\epsilon(\theta)}\left({L_n}(c)-L_n(\theta)\right) -\limsup_{n\rightarrow \infty}\sup_{c\not\in  B_\epsilon(\theta)}\left({L_n}(c)-L_n(\theta)\right)\\
\leq\eta=-\frac{1}{2} \limsup_{n\rightarrow\infty}\sup_{c\not\in B_\epsilon(\theta)}\left({L_n}(c)-L_n(\theta)\right),
\end{multline*}
which implies $\frac{1}{2} \limsup_{n\rightarrow\infty}\sup_{c\not\in B_\epsilon(\theta)}\left({L_n}(c)-L_n(\theta)\right) \geq \sup_{c\not\in  B_\epsilon(\theta)}\left({L_n}(c)-L_n(\theta)\right)$.

Therefore, w.p.a.1, 
\[
L_n(\hat{\theta})-L_n({\theta})>\sup_{c\not\in  B_\epsilon(\theta)}\left({L_n}(c)-L_n(\theta)\right),
\]
which implies that $\hat{\theta}\in B_\epsilon(\theta)$ w.p.a.1. Because $\epsilon$ can be arbitrarily small, $\widehat{\theta}\overset{p}{\rightarrow}\theta$.
\end{proof}
\end{lemma}

\begin{lemma}

\label{lemma3}
Suppose that \cref{extremevalue}, \ref{compact}--(i) and \ref{stable2} hold. Then, 
\[
\hat{L}(\hat{\theta})\geq \sup_{c\in\Theta}\hat{L}(c)-o_p(1).
\]
\proof 
By the definition of $\hat\theta$, it suffices to show that $\sup_{c\in\Theta}\big|\hat{Q}(c)-\hat{L}(c)\big|\rightarrow 0$. 

Because
\[
\sup_{c\in\Theta}|\hat{Q}(c)-\hat{L}(c)|
\leq \sup_{c\in\Theta}\frac{1}{n}\sum_{i=1}^n\sum_{k\in A}\big|\ln\sigma^h_{ik}\left(\mathbb S; c\right)-\ln \sigma^*_{ik}(\mathbb S; c)\big|.
\]
By Taylor expansion, 
\[
\sum_{k\in A}\big|\ln\sigma^h_{ik}\left(\mathbb S |c\right)-\ln \sigma^*_{ik}(\mathbb S; c)\big|=\frac{1}{\sigma^\dag}\sum_{k\in A}\big|\sigma^h_{ik}\left(\mathbb S; c\right)- \sigma^*_{ik}(\mathbb S; c)\big|\leq \frac{2\lambda^{h+1}}{{\sigma}_0},
\]
where ${\sigma}^\dag$ is some real value between $\sigma^h_{ik}\left(\mathbb S; c\right)$ and $\sigma^*_{ik}(\mathbb S; c)$, and $\sigma_0$ is the lower bound of the equilibrium choice probability. The last step uses \Cref{bounded,stability}. Thus,
\[
\sup_{c\in \Theta}\big|\hat Q(c)-\hat{L}(c)\big|\leq \frac{2\lambda^{h+1}}{{\sigma}_0}.
\]
Because of \cref{choiceh} and $\lambda<1$, we have $\sup_{c\in\Theta}\big|\hat Q(c)-\hat{L}(c)\big|\overset{p}{\rightarrow} 0$.
\qed
\end{lemma}

\subsection{Proof of Theorem \ref{normality}}
\label{proofnormality}
\proof 
First, by the proof of \Cref{lemma3} and \cref{choiceh}--(ii),
\[
\sup_{c\in \Theta}\big|\hat Q(c)-\hat{L}(c)\big|\leq \frac{2(K+1)\lambda^{h}}{{\sigma}_0}=o_p(n^{-1}).
\]Hence, $\hat L(\hat\theta)\geq \sup_{c\in\Theta} \hat L(c)-o_p(n^{-1})$, which implies that $
{\partial \hat L(\hat\theta)}/{\partial c}=o_p(n^{-1/2})$.

By the Taylor expansion, we have
\begin{equation*}
\frac{\partial \hat L(\theta)}{\partial c}+\frac{\partial^2 \hat{L}({\theta}^\dag)}{\partial c\partial c'}(\hat{\theta}-\theta)=o_p(n^{-1/2})
\end{equation*}for some $\theta^\dag$ between $\theta$ and $\hat{\theta}$. Now it suffices to show: 
\begin{align}
\label{condition1}
&\sqrt{n}\times \frac{\partial \hat L(\theta)}{\partial c}\overset{d}{\rightarrow}  N\big (0, J(\theta)\big),\\
\label{condition2}
&\frac{\partial^2 \hat{L}({\theta}^\dag)}{\partial c\partial c'}\overset{p}{\rightarrow}-J(\theta).
\end{align}

We first show \cref{condition1}. Let $\xi_{i}=\frac{\partial}{\partial c} \sum_{k\in A}\one(Y_i=k)\ln \sigma^*_{ik}(\mathbb S; c)|_{c=\theta}$. Note that the true parameter $\theta$ always maximizes the likelihood function $\mathbb{E}\big[\sum_{k\in A}\one(Y_i=k)\ln \sigma^*_{ik}(\mathbb S; \cdot)|\mathbb S\big]$ for any $n$ and $\mathbb S$. Thus $\mathbb{E}\left(\xi_i|\mathbb S\right)=0$. 

By definition, $\partial \hat L(\theta)/\partial c=n^{-1}\sum_{i=1}^n\xi_i$. Then, it suffices to show that $n^{-1/2} \sum_{i=1}^n \xi_i\overset{d}{\rightarrow} N(0, J(\theta)) $. Equivalently, we need to show $ n^{-1/2} \sum_{i=1}^n J(\theta)^{-\frac{1}{2}}\xi_i\overset{d}{\rightarrow} N(0, \textbf 1_{P}) $, where $\textbf{1}_P$ is the $P$--by--$P$ identity matrix. For this, we show that the conditional distribution of $\sqrt n \sum_{i=1}^n J(\theta)^{-\frac{1}{2}}\xi_i$ given $\mathbb S$ always converges to the same limiting normal distribution $N(0, \textbf 1_{P})$. 

Because $\xi_i$ is conditionally independent across $i$ given $\mathbb S$. Then
\[
\mathbb{E}\left[\big(n^{-1/2}\sum_{i=1}^n\xi_i\big)\cdot\big(n^{-1/2}\sum_{i=1}^n \xi_i'\big)\Big|\mathbb S\right]=n^{-1}\sum_{i=1}^n\mathbb{E}\left(\xi_i\cdot\xi_i'\big|\mathbb S\right).
\]By a similar argument to that in the proof of \Cref{asymptotics}, we have 
\[
n^{-1}\sum_{i=1}^n\mathbb{E}\left(\xi_i\cdot\xi_i'\big|\mathbb S\right)=n^{-1}\sum_{i=1}^n\mathbb{E}\left(\xi_i\cdot\xi_i'\right)+o_p(1)=J_n(\theta)+o_p(1)=J(\theta)+o_p(1).
\]Thus, 
\[
\mathbb{E}\left[\big(n^{-1/2}\sum_{i=1}^n\xi_i\big)\cdot\big(n^{-1/2}\sum_{i=1}^n \xi_i'\big)\Big|\mathbb S\right]\overset{p}{\rightarrow} J(\theta).
\]
Hence, by the Lindeberg-Feller Theorem  \citep[see e.g.][]{van2000asymptotic}, conditional on $\mathbb S$,
 \[
 n^{-1/2} \sum_{i=1}^n J(\theta)^{-\frac{1}{2}}\xi_i\overset{d}{\rightarrow} N(0, \textbf 1_{P})
 \]

We now show \cref{condition2}. Under \cref{compact}, it follows from \Cref{bounded,c5} that $\left\|\frac{\partial^2}{\partial c\partial c'} \sum_{k\in A} \one(Y_i=k)\ln \sigma^*_{ik}(\mathbb S; c)\right\|$  is bounded above uniformly on $n$, $\mathbb S$ and $\theta$, and $\frac{\partial^2}{\partial c\partial c'} \sum_{k\in A} \one(Y_i=k)\ln \sigma^*_{ik}(\mathbb S; c)$ are smooth functions of $c\in\Theta$. Hence by a similar argument as the proofs in  \Cref{asymptotics},
\[
\sup_{c\in\Theta}\left[\frac{\partial \hat{L}(c)}{\partial c\partial c'} -\frac{1}{n}\sum_{i=1}^n\mathbb{E}\left\{\frac{\partial^2}{\partial c\partial c'} \sum_{k\in A}\one(Y_i=k)\ln \sigma^*_{ik}(\mathbb S; c)\right\} \right]\overset{p}{\rightarrow}0.
\]
Because $\theta^\dag\overset{p}{\rightarrow}\theta$ and by \cref{SNdistr},  we have 
\[
\frac{\partial^2\hat{L}(\theta^\dag)}{\partial c\partial c'} =\mathbb{E}\left\{\frac{\partial^2}{\partial c\partial c'} \sum_{k\in A}\one(Y_1=k)\ln \sigma^*_{1k}(\mathbb S; \theta)\right\}+o_p(1).
\]
Moreover, by the information matrix equality, 
\[
\mathbb{E}\left\{\frac{\partial^2}{\partial c\partial c'} \sum_{k\in A}\one(Y_1=k)\ln \sigma^*_{1k}(\mathbb S; \theta)\right\}=-J_n(\theta)=-J(\theta)+o(1).
\] Then \cref{condition2} is proved.
\qed

\section{Auxiliary lemmas}

\begin{lemma}
\label{bounded}
Suppose \cref{extremevalue} and \ref{compact}--(i) hold. Then there exists $\sigma_0\in(0,1)$ such that 
$ \sigma^*_{ik}(\mathbb S; c)\geq{\sigma}_0$ for all $n\in\mathbb N$, $i\in N$, $k\in A$ and $c\in\Theta$.

\proof  By \cref{extremevalue}, for all $(i,k)\in N\times A$,
\begin{equation*}
 \sigma^*_{ik}(\mathbb S; c)=\frac{\exp\left\{(X'_i,Q_i)\cdot b_k+\sum_{\ell\in A}a_{k\ell}\big(\frac{1}{Q_i}\sum_{j\in F_i}\sigma^*_{j\ell}(\mathbb S; c)\big)\right\}}{1+\sum_{\ell'=1}^K\exp\left\{(X'_i,Q_i)b_{\ell'}+\sum_{\ell\in A}a_{\ell'\ell}\big(\frac{1}{Q_i}\sum_{j\in F_i}\sigma^*_{j\ell'}(\mathbb S; c)\big)\right\}}.
\end{equation*} 
Because $0\leq \frac{1}{Q_i}\sum_{j\in F_i}\sigma^*_{j\ell}(\mathbb S; c)\leq 1$ and by \cref{compact}--(i),  the RHS has a lower bound, denoted as $\sigma_0>0$. Note that the above argument does not depend on the value of $n$, $i$, $k$ and $c$.  
\qed
\end{lemma}

\begin{lemma}
\label{c5}
Suppose that \cref{extremevalue,stable2}  hold. Then, $\sigma^*_{ik}(\mathbb S; \cdot)\in\mathcal{C}^\infty(\Theta)$ for all $n\in \mathbb N$, $\mathbb S$, $i\in N$, $k\in A$ and $c\in\Theta$.

\proof  We fix an arbitrary $n$ and $\mathbb S$ in the following analysis. By \Cref{uniqueness}, $\{\sigma^*_{i}(\mathbb S; c): i\in N\}$ is the unique solution to the equation system: for all $(i,k)\in (N,A)$,
\[
\sigma^*_{ik}=\frac{\exp\left[b_k(X_i,Q_i)+\sum_{l=0}^K\big\{a_k(\ell,X_i,Q_i)\cdot\sum_{j\in F_i}\sigma^*_{j\ell}\big\}\right]}{1+\sum_{q=1}^K\exp\left[b_q(X_i,Q_i)+\sum_{l=0}^K\big\{a_q(\ell,X_i,Q_i)\cdot\sum_{j\in  F_i}\sigma^*_{j\ell}\big\}\right]}.
\] Let $\Sigma^*=({\sigma_1^*},\cdots,{\sigma^*_n})$. Then the above equation system can be represented as
\[
\Sigma^*= BR(\mathbb S, \Sigma^*  ; c)
\]where $BR$ is the $n(K+1)$ dimensional mapping representing the best response functions for all $(i,k)\in (N,A)$.  Fix $\mathbb S$. Clearly, BR belongs to $ \mathcal{C}^{\infty}\left(\mathbb{R}^{n(K+1)}\times\Theta\right)$.  Then by implicit function theorem, the solution $\sigma^*_{i}(\mathbb S;\cdot)\in\mathcal{C}^\infty\left(\Theta\right)$ for all $i\in N$.
\qed
\end{lemma}

\section{Consistent nonparametric estimator of $\mathbb P_{Y_i|\mathbb S}$}
\label{appendixd}
The NDD condition is important for large network asymptotics. In particular, it allows us to nonparametrically estimate the probability distribution  $\mathbb P_{Y_i|\mathbb S}$ using observations from one single large network. To illustrate,  we consider the simple circle network where each player has two direct friends and the friendship is symmetric. Such a specification helps highlights key features of the consistency argument for  the nonparametric estimation.

Because our asymptotic analysis considers a sequence of games with $n\rightarrow \infty$,  we use $\mathbb S_n$ with  subscript $n$ to emphasize its dependence on the network size in the following analysis. The sequence of games are described as follows: Let the set of players $\{1,2,\cdots,n\}$ for $n\geq 2$ be located on a circle network as follows: First we randomly pick a location for player 1 on the circle. Next, players 2 and 3 are on 1's left and right, respectively; then players 4 and 5 are further located on 2's left and 3's right, respectively; so on and so forth. Thus we obtain a circle network with $n=+\infty$ in the limit. Given the network, state variables $X_i$ are i.i.d. across all the players. Similarly to the probability theory in time series, the probability distribution of the sequence $\{\mathbb S_n: n\geq 2\}$ is well defined.

For simplicity, let $\mathbb A=\{0,1\}$ and $X_i\in\mathbb R$. W.o.l.g., we consider the estimation of $\mathbb P(Y_i=1|\mathbb S_n=s_n)$ for $i=1$. To begin with, we first consider the case where $X_i$ is binary, i.e., $X_i\in\{0,1\}$. It is straightforward that our arguments can be generalized to the case of multiple valued $X_i$'s. The continuous $X_i$'s case will be discussed later. Intuitively, a nonparametric estimator $\hat{\mathbb P} (Y_1=1|\mathbb S_n=s_n)$ can be defined as follows:
\[
\frac{\sum_{j=1}^n \one (Y_j=1)\cdot  \mathbb 1 \left[\mathbb G_{(j,h)}= g_{(1,h)}\right]\cdot \mathbb 1\left[ X_{j(\ell)}=x_{1(\ell)}, \ \ \text{for }\ell =-h,\cdots,h\right]}{\sum_{j=1}^n  \mathbb 1 \left[\mathbb G_{(j,h)}= g_{(1,h)}\right]\cdot \mathbb 1\left[ X_{j(\ell)}=x_{1(\ell)}, \ \ \text{for }\ell =-h,\cdots,h\right]},
\]where $j(\ell)$ denotes the $|\ell|$-th left vertex of $j$ if $\ell<0$; otherwise it refers to the $|\ell|$-th right vertex of $j$. Note that because of the circle network, $\mathbb G_{(j,h)}= g_{(1,h)}$ a.s.. Then, the term $ \mathbb 1 \left[\mathbb G_{(j,h)}= g_{(1,h)}\right]$ is redundant in the above expression. As is shown in the proof of the next lemma, the above estimator is essentially a kernel estimator with a specific choice of bandwidth and a uniform kernel. 

In the above estimator, It is crucial to choose $h$ for its consistency, which carries a bias and variance tradeoff: Intuitively, $h\in\mathbb N$ needs to increase properly with $n$ such that $\mathbb P(Y_1=1|\mathbb S_{(1,h)})$ converges to $\mathbb P(Y_1=1|\mathbb S_n)$ (note that the approximation error is bounded by $2\xi^{h+1}$ where $|\xi|<1$). On the other hand, we require the number of observations $\mathbb G_{(j,h)}= g_{(1,h)}$ goes to infinity with the network size, so that the variance of the estimator decreases to zero as $n\rightarrow \infty$. 

W.l.o.g., suppose $\mathbb P(X_i=0)\leq 1/2$. Moreover, let $p_h\equiv \mathbb P(\mathbb S_{(1,h)}=s_{(1,h)})=\prod_{j=1}^{2h+1} \mathbb P (X_j=x_j)$. By definition, $\mathbb P(X_i=0)^{2h+1}\leq p_h\leq \mathbb P(X_i=1)^{2h+1}$. Therefore, we have $p_h\rightarrow 0$ as $h\rightarrow \infty$. 
\begin{lemma}
\label{nconsistency}
Suppose that  \cref{extremevalue,para-id}, \ref{compact}-(i), \ref{SNdistr} and \ref{maximumneighbours} hold. Suppose $h\rightarrow \infty$ and $\frac{h}{np_h}\rightarrow 0$ as $n\rightarrow \infty$. Then
\[
\hat{\mathbb P} (Y_1=1|\mathbb S_n=s_n)-\mathbb P (Y_1=1|\mathbb S_n=s_n)\overset{p}{\rightarrow}0
\]
\proof
First note that 
\begin{multline*}
\hat{\mathbb P} (Y_1=1|\mathbb S_n=s_n)\\
=\frac{\frac{1}{np_h}\sum_{j=1}^n \one (Y_j=1)\cdot  \mathbb 1 \left[\mathbb G_{(j,h)}= g_{(1,h)}\right]\cdot \mathbb 1\left[ X_{j(\ell)}=x_{1(\ell)}, \ \ \text{for }\ell =-h,\cdots,h\right]}{\frac{1}{np_h}\sum_{j=1}^n  \mathbb 1 \left[\mathbb G_{(j,h)}= g_{(1,h)}\right]\cdot \mathbb 1\left[ X_{j(\ell)}=x_{1(\ell)}, \ \ \text{for }\ell =-h,\cdots,h\right]},
\end{multline*}
We now show that the denominator  and   numerator converge to one and  $\mathbb P(Y_1=1|\mathbb S_{(1,h)}=s_{(1,h)})$, respectively. First, we  look at denominator and show 
\begin{align}
\label{nbias}
&\mathbb E\left\{\frac{1}{np_h}\sum_{j=1}^n  \mathbb 1 \left[\mathbb G_{(j,h)}= g_{(1,h)}\right]\cdot \mathbb 1\left[ X_{j(\ell)}=x_{1(\ell)}, \ \ \text{for }\ell =-h,\cdots,h\right]\right\}\rightarrow 1;\\
\label{nvar}
&\text{Var}\left\{\frac{1}{np_h}\sum_{j=1}^n  \mathbb 1 \left[\mathbb G_{(j,h)}= g_{(1,h)}\right]\cdot \mathbb 1\left[ X_{j(\ell)}=x_{1(\ell)}, \ \ \text{for }\ell =-h,\cdots,h\right]\right\}\rightarrow 0.
\end{align} Regarding \eqref{nbias}, we have
\begin{multline*}
\mathbb E\left\{\frac{1}{np_h}\sum_{j=1}^n  \mathbb 1 \left[\mathbb G_{(j,h)}= g_{(1,h)}\right]\cdot  \mathbb 1\left[\{ X_\ell: \ell\in N_{(j,h)}\}=\{x_\ell: \ell \in N_{(1,h)}\}\right]\right\}\\
=\frac{1}{p_h}\mathbb E\left\{  \mathbb 1 \big[\mathbb S_{(1,h)}=s_{(1,h)}\big]\right\}=1.
\end{multline*}
To establish \eqref{nvar}, note that
\begin{eqnarray*}
&&\text{Var}\left\{\frac{1}{np_h}\sum_{j=1}^n  \mathbb 1 \left[\mathbb G_{(j,h)}= g_{(1,h)}\right]\cdot  \mathbb 1\left[\{ X_\ell: \ell\in N_{(j,h)}\}=\{x_\ell: \ell \in N_{(1,h)}\}\right]\right\}\\
&=&\frac{1}{n^2p^2_h}\sum_{\ell=1}^n \sum_{j\neq \ell}\text {Cov}\Big\{  \mathbb 1 \big[\mathbb S_{(j,h)}= s_{(1,h)}\big], \mathbb 1 \big[\mathbb S_{(\ell,h)}= s_{(1,h)}\big]\Big\}+\frac{1}{np^2_h}\text {Var}\Big\{  \mathbb 1 \big[\mathbb S_{(1,h)}= s_{(1,h)}\big]\Big\}\\
&=&\frac{1}{np^2_h}\sum_{j\neq 1} \text {Cov}\Big\{  \mathbb 1 \big[\mathbb S_{(j,h)}= s_{(1,h)}\big], \mathbb 1 \big[\mathbb S_{(1,h)}= s_{(1,h)}\big]\Big\}+\frac{1-p_h}{np_h}\\
&=&\frac{1}{np^2_h}\sum_{j=2}^{2h+1} \text {Cov}\Big\{  \mathbb 1 \big[\mathbb S_{(j,h)}= s_{(1,h)}\big], \mathbb 1 \big[\mathbb S_{(1,h)}= s_{(1,h)}\big]\Big\}+\frac{1-p_h}{np_h},
\end{eqnarray*}where the last step comes from the assumption that $S_{(j,h)}$ is independent of $S_{(1,h)}$ if $N_{(j,h)}$ does not overlap with $N_{(1,h)}$. Thus, 
\begin{eqnarray*}
&&\text{Var}\left\{\frac{1}{np_h}\sum_{j=1}^n  \mathbb 1 \left[\mathbb G_{(j,h)}= g_{(1,h)}\right]\cdot  \mathbb 1\left[\{ X_\ell: \ell\in N_{(j,h)}\}=\{x_\ell: \ell \in N_{(1,h)}\}\right]\right\}\\
&\leq &\frac{2h}{np^2_h}\times \frac{\text {Var}\Big\{  \mathbb 1 \big[\mathbb S_{(j,h)}= s_{(1,h)}\big]\Big\}+\text {Var}\Big\{  \mathbb 1 \big[\mathbb S_{(1,h)}= s_{(1,h)}\big]\Big\}}{2}+ \frac{1-p_h}{np_h} \\
&=&\frac{(2h+1)(1-p_h)}{np_h}\propto \frac{h}{np_h}\rightarrow 0.
\end{eqnarray*}
It follows that 
\[
\frac{1}{np_h}\sum_{j=1}^n  \mathbb 1 \left[\mathbb G_{(j,h)}= g_{(1,h)}\right]\cdot \mathbb 1\left[ X_{j(\ell)}=x_{1(\ell)}, \ \ \text{for }\ell =-h,\cdots,h\right]\overset{p}{\rightarrow} 1
\]

By a similar argument, we have
\begin{eqnarray*}
&&\mathbb E \left\{\frac{1}{np_h}\sum_{j=1}^n  \mathbb 1(Y_j=1)\cdot \mathbb 1 \left[\mathbb G_{(j,h)}= g_{(1,h)}\right]\cdot \mathbb 1\left[ X_{j(\ell)}=x_{1(\ell)}, \ \ \text{for }\ell =-h,\cdots,h\right]\right\}\\
&=&\mathbb P (Y_1=1|\mathbb S_{(1,h)}=s_{(1,h)})=\mathbb P (Y_1=k|\mathbb S_n=s_n)+o(|\xi|^h)
\end{eqnarray*}and 
\[
\text{Var}\left\{\frac{1}{np_h}\sum_{j=1}^n \mathbb 1(Y_j=1)\cdot  \mathbb 1 \left[\mathbb G_{(j,h)}= g_{(1,h)}\right]\cdot \mathbb 1\left[ X_{j(\ell)}=x_{1(\ell)}, \ \ \text{for }\ell =-h,\cdots,h\right]\right\}\rightarrow 0.
\]
Moreover, by Slutsky's theorem, we establish the consistency of the proposed estimator.\qed
\end{lemma}
\noindent
In \Cref{nconsistency}, it is required that $h$ should increase to infinity with $n$, but sufficiently slow. In particular, the conditions imply $p_h\rightarrow 0$ and $np_h\rightarrow \infty$ as $n\rightarrow \infty$. This suggests that the term $p_h$ plays the same role as the bandwidth in kernel estimation. In addition,  because of the dependence between $S_{(j,h)}$ and $S_{(i,h)}$ for $\rho(i,j)\leq h$,  we require $np_h$ increase to infinity faster than $h$. Suppose one chooses $h=[h_0\times \ln n]$ for some constant $h_0>0$. Then, $p_h\propto n^{-\kappa}$ where $\kappa>0$ that is determined by $h_0$ and $\mathbb P(X_i=0)$. Then the restrictions on $h$ in \Cref{nconsistency} are satisfied if $\kappa$ is sufficiently small.

Suppose $X_i$ is continuously distributed. Let  $f_X$ be the pdf of $X_i$. For simplicity, We assume $0<\inf_{x\in\mathbb R} f_X(x)<\sup_{x\in \mathbb R} f_X(x)<\infty$. As usual,  additional assumptions on the structural parameters are needed to ensure $\mathbb P(Y_1=1|\mathbb S_n=s_n)$ is $R$--th ($R\geq 2$) order continuously differentiable in each argument of $\mathbb S_n$. Moreover, a nonparametric estimator is defined by
\[
\hat{\mathbb P} (Y_1=1|\mathbb S_n=s_n)=\frac{\sum_{j=1}^n \one (Y_j=1)\cdot  \mathbb 1 \left[\mathbb G_{(j,h)}= g_{(1,h)}\right]\cdot \prod_{\ell=1}^{2h+1} K\left(\frac{X_{\ell}-x_\ell}{b_\ell}\right)}{\sum_{j=1}^n  \mathbb 1 \left[\mathbb G_{(j,h)}= g_{(1,h)}\right]\cdot \prod_{\ell=1}^{2h+1} K\left(\frac{X_{\ell}-x_\ell}{b_\ell}\right)},
\]where $K$ and $b_\ell$ for $\ell=1,\cdots,2h+1$ are $R$--th order kernel function and  bandwidth, respectively. 


For consistency, we need to choose $h\rightarrow \infty$ and  $b_\ell\rightarrow 0$ for $\ell=1,\cdots,2h+1$ properly as $n\rightarrow\infty$. For simplicity, let $b_\ell f_X(x_\ell)=p$ for some $p\equiv p_n>0$. Moreover, let $h\rightarrow \infty$, $p\rightarrow 0$ and $h/(np^{2h+1})\rightarrow 0$ as $n\rightarrow \infty$.  By a similar argument to \Cref{nconsistency} and Bochner's Lemma, we can show consistency of the kernel estimator. In particular, we have
\[
\mathbb E \left\{\frac{1}{np^{2h+1}}\sum_{j=1}^n  \mathbb 1 \left[\mathbb G_{(j,h)}= g_{(1,h)}\right]\cdot \prod_{\ell=1}^{2h+1} K\left(\frac{X_{\ell}-x_\ell}{b_\ell}\right)\right\}= 1+ O(p^R)
\] and
\[
\text{Var} \left\{\frac{1}{np^{2h+1}}\sum_{j=1}^n  \mathbb 1 \left[\mathbb G_{(j,h)}= g_{(1,h)}\right]\cdot \prod_{\ell=1}^{2h+1} K\left(\frac{X_{\ell}-x_\ell}{b_\ell}\right)\right\}= O\big(\frac{h}{n p^{2h+1}}\big),
\]and similar expressions hold for the numerator of the kernel estimator, which provide the consistency.

\end{document}